\documentclass[acmsmall,nonacm]{acmart}
\usepackage{paper-authors}

\setcopyright{cc}
\setcctype{by}
\acmDOI{}
\usepackage{graphicx}
\usepackage{wrapfig}
\usepackage{xcolor}
\usepackage{amsmath}
\usepackage{amsfonts}

\usepackage{amssymb}
\usepackage{mathtools}
\usepackage{stmaryrd}
\usepackage[inline]{enumitem}
\usepackage{thm-restate}
\usepackage{tikz}
\usetikzlibrary{arrows,arrows.meta,automata,positioning,decorations.pathreplacing,calligraphy}
\usepackage{algorithm}
\usepackage{algpseudocode}
\algtext*{EndFor}
\usepackage{soul}
\usetikzlibrary{tikzmark,calc}
\usepackage{bbm}

\usepackage{listings}

\allowdisplaybreaks[3]
\makeatletter
\g@addto@macro\@adjustvertspacing{%
  \setlength{\abovedisplayskip}{2pt plus 1pt minus 0.5pt}%
  \setlength{\belowdisplayskip}{2pt plus 1pt minus 0.5pt}%
  \setlength{\abovedisplayshortskip}{0pt plus 1pt}%
  \setlength{\belowdisplayshortskip}{2pt plus 1pt minus 0.5pt}%
  \setlength{\jot}{1pt}%
}
\makeatother
\AtBeginDocument{%
  \setlength{\floatsep}{8pt plus 2pt minus 2pt}%
  \setlength{\textfloatsep}{10pt plus 2pt minus 2pt}%
  \setlength{\intextsep}{8pt plus 2pt minus 2pt}%
  \setlength{\dblfloatsep}{8pt plus 2pt minus 2pt}%
  \setlength{\dbltextfloatsep}{10pt plus 2pt minus 2pt}%
}

\makeatletter
\expandafter\patchcmd\csname\string\proof\endcsname{\topsep6\p@\@plus6\p@\relax}{\topsep2\p@\@plus1\p@\relax}{}{\PackageError{paper-layout}{Proof spacing patch failed}{}}
\makeatother

\newenvironment{myparagraph}[1]{\par\addvspace{2pt plus 1pt minus 0.5pt}\noindent{\bf #1.}}{}

\definecolor{literalInk}{HTML}{75415F}
\DeclareRobustCommand{\strlit}[1]{%
  \ifmmode
    \text{\normalfont\ttfamily\bfseries\color{literalInk}#1}%
  \else
    {\normalfont\ttfamily\bfseries\color{literalInk}#1}%
  \fi}

\newcommand{\leanchecked}{}
\theoremstyle{definition}
\newtheorem{theorem}{Theorem}
\newtheorem{definition}[theorem]{Definition}
\newtheorem{corollary}[theorem]{Corollary}
\newtheorem{lemma}[theorem]{Lemma}
\newtheorem{proposition}[theorem]{Proposition}

\newcommand{\figlabel}[1]{\label{fig:#1}}
\newcommand{\figref}[1]{Fig.~\ref{fig:#1}}
\newcommand{\seclabel}[1]{\label{sec:#1}}
\newcommand{\secref}[1]{Section~\ref{sec:#1}}

\newcommand{\deflabel}[1]{\label{def:#1}}
\newcommand{\defref}[1]{Definition~\ref{def:#1}}
\newcommand{\problabel}[1]{\label{prob:#1}}
\newcommand{\probref}[1]{Problem~\ref{prob:#1}}
\newcommand{\thmlabel}[1]{\label{thm:#1}}
\newcommand{\thmref}[1]{Theorem~\ref{thm:#1}}
\newcommand{\proplabel}[1]{\label{prop:#1}}
\newcommand{\propref}[1]{Proposition~\ref{prop:#1}}
\newcommand{\lemlabel}[1]{\label{lem:#1}}
\newcommand{\lemref}[1]{Lemma~\ref{lem:#1}}
\newcommand{\corlabel}[1]{\label{cor:#1}}
\newcommand{\corref}[1]{Corollary~\ref{cor:#1}}

\newcommand{\equlabel}[1]{\label{eq:#1}}

\newcommand{\calA}{\mathcal{A}}

\newcommand{\calD}{\mathcal{D}}
\newcommand{\calF}{\mathcal{F}}
\newcommand{\calG}{\mathcal{G}}
\newcommand{\calL}{\mathcal{L}}
\newcommand{\calH}{\mathcal{H}}

\newcommand{\calS}{\mathcal{S}}

\renewcommand{\emptyset}{\varnothing}
\newcommand{\set}[1]{\{#1\}}
\newcommand{\setpred}[2]{\{#1 \,|\, #2\}}

\newcommand{\norm}[1]{\left| #1 \right|}

\newcommand{\nats}{\mathbb{N}}

\newcommand{\poly}[1]{\textsf{poly}(#1)}
\newcommand{\sharpeP}{\#\textsf{P}}
\newcommand{\cntNFA}{\ensuremath{\mathsf{\#NFA}}}
\newcommand{\cntSNFA}{\ensuremath{\mathsf{\#Symbolic\text{-}NFA}}}

\newcommand{\emptyword}{\lambda}
\newcommand{\len}[1]{|#1|}
\newcommand{\concat}{\cdot}

\newcommand{\aut}{\calA}
\newcommand{\saut}{\calA}
\newcommand{\states}{Q}

\newcommand{\qI}{q_{\mathsf{I}}}
\newcommand{\qF}{q_{\mathsf{F}}}
\newcommand{\accept}{F}
\newcommand{\trans}{\Delta}
\newcommand{\lang}[1]{\calL(#1)}
\newcommand{\langstate}[1]{L(#1)}
\newcommand{\intrans}[1]{\Delta_{\mathsf{in}}(#1)}

\newcommand{\unroll}{\textsf{u}}

\newcommand{\sautunroll}{\saut^\unroll}
\newcommand{\statesunroll}{\states^\unroll}
\newcommand{\transunroll}{\trans^\unroll}

\newcommand{\domain}{\mathbb{D}}
\newcommand{\predset}{\Psi}
\newcommand{\predphi}{\varphi}
\newcommand{\predpsi}{\psi}
\newcommand{\denote}[1]{\llbracket #1 \rrbracket}
\newcommand{\sat}{\models}

\newcommand{\cnt}[1]{\#(#1)}

\newcommand{\numwords}[2]{\#_{#2}(#1)}

\newcommand{\epsl}{\epsilon}
\newcommand{\dl}{\delta}
\newcommand{\Ex}{\mathsf{E}}

\newcommand{\core}{\mathsf{Core}}
\newcommand{\median}{\mathsf{median}}
\newcommand{\minsf}{\mathsf{min}}

\newcommand{\approxalg}{\mathsf{ApproxSymNFA}}
\newcommand{\estsamp}{\mathsf{estimateAndSample}}
\newcommand{\corestar}{\core^{*}}
\newcommand{\corespirit}{\core^{\mathsf{s}}}
\newcommand{\acnt}[1]{\mathsf{c}(#1)}          % memoized approximate count
\newcommand{\osample}[1]{\mathsf{sample}(#1)}  % oracle: almost-uniform sampler
\newcommand{\omember}{\mathsf{member}}         % oracle: membership test
\newcommand{\sdist}[1]{\pi_{#1}}               % sampler's actual distribution
\newcommand{\dtol}{\kappa}                     % oracle tolerance
\newcommand{\Sbar}{\bar{S}}
\newcommand{\Shat}{\hat{S}}
\newcommand{\outtrans}[1]{\Delta_{\mathsf{out}}(#1)}
\newcommand{\eos}{\rhd}                        % fresh end-marker letter
\newcommand{\canon}{\mathsf{ct}}               % canonical transition
\newcommand{\cpath}[2]{\mathsf{path}(#1,#2)}   % canonical run
\newcommand{\Tmem}{T_{\mathsf{mem}}}
\newcommand{\Tsamp}{T_{\mathsf{samp}}}
\newcommand{\Tcnt}{T_{\mathsf{cnt}}}
\newcommand{\thetas}{\theta_{\mathsf{s}}}       % sampling budget
\newcommand{\thetar}{\theta_{\mathsf{r}}}       % retention budget
\newcommand{\ctrs}{\mathit{ctr}_{\mathsf{s}}}   % sampling counter
\newcommand{\ctrr}{\mathit{ctr}_{\mathsf{r}}}   % retention counter
\newcommand{\dmax}{d_{\mathsf{in}}}             % max in-degree

\newcommand{\mass}[1]{\mathsf{M}(#1)}          % weighted mass of L(q)
\newcommand{\wgt}{\Lambda}                     % per-word weight
\newcommand{\tolm}{\tau}                       % graded lower-slack
\newcommand{\driftXi}{\xi}                     % drift bound
\newcommand{\divv}[2]{q^{#1,#2}}               % divergence state
\newcommand{\Div}{\mathsf{div}}                % divergence level
\newcommand{\Dset}[2]{D(#1,#2)}                % divergence classes
\newcommand{\ind}[1]{\mathbbm{1}_{#1}}
\newcommand{\cache}{\mathsf{cache}}

\definecolor{lightyellow}{rgb}{1,1,0.7}

\begin{document}

\title{Approximate Counting for Symbolic Automata}
\titlenote{Full version of the paper accepted at POPL 2027.}
\titlenote{All authors contributed equally to this research. The symbol \textcircled{r} denotes random author order. The publicly verifiable record of the randomization is available at \protect \url{https://www.aeaweb.org/journals/policies/random-author-order} with confirmation code: {oRWFAcu-XdGt}}

\author{Kuldeep S. Meel}
\orcid{0000-0001-9423-5270}
\affiliation{%
  \institution{University of Toronto}
  \city{Toronto}
  \country{Canada}
}
\affiliation{%
  \institution{Georgia Institute of Technology}
  \city{Atlanta}
  \country{USA}
}
\email{meel@cs.toronto.edu}

\author{Umang Mathur}
\correspondingauthor
\orcid{0000-0002-7610-0660}
\affiliation{%
  \institution{National University of Singapore}
  \department{School of Computing}
  \city{Singapore}
  \country{Singapore}
}
\email{umathur@nus.edu.sg}

\author{Alexis de Colnet}
\orcid{0000-0002-7517-6735}
\affiliation{%
  \institution{Leiden University}
  \city{Leiden}
  \country{Netherlands}
}
\email{decolnet@ac.tuwien.ac.at}

%!TEX root=./main.tex

\begin{abstract}
  Given a nondeterministic finite automaton $\aut$ and a
  length $n$ in unary, the problem \cntNFA{} seeks to compute the
  number of words of length $n$ accepted by $\aut$.
  A celebrated result of Arenas, Croquevielle, Jayaram, and Riveros established
  that \cntNFA{} admits a fully polynomial-time randomized
  approximation scheme (FPRAS).
  Subsequent works have brought the running time down to
  $O(n^2m^3\log(nm)\,\epsl^{-2}\log(1/\dl))$ for an NFA with $m$ states
  over a binary alphabet, with relative error at most $\epsl$ and failure
  probability at most $\dl$.
  These running-time bounds concern explicitly represented alphabets,
  whereas automata arising from programs, strings, and structured data are naturally \emph{symbolic},
  i.e., their transitions carry predicates, or guards, over a large
  concrete alphabet $\domain$.

  In this work, we study \cntSNFA{}, which asks to count the number of words
  of a given length accepted by a symbolic NFA. We present a randomized
  approximation algorithm for this $\sharpeP$-hard problem.
  We formulate the algorithm using an oracle
  interface that supports membership testing, approximate counting, and
  almost-uniform sampling for transition guards. With polynomial-time
  implementations of these guard operations, the algorithm yields an FPRAS.
  Our algorithm extends the recent dependent sampling approach of Meel and
  de Colnet to symbolic automata. Our FPRAS queries only the input guards,
  avoiding the potentially exponential blowup from constructing Boolean
  combinations. For an automaton with transition set $\trans$, the algorithm makes
  $O(n^2\norm{\trans}\log(n\norm{\trans})\,\epsl^{-2}\log(1/\dl))$ counting and
  sampling queries. For constant-size explicit alphabets, our algorithm
  recovers the above running-time bound for \cntNFA{}.
  We also formalize the core probabilistic analysis, including concentration
  and confidence amplification, in Lean 4.
\end{abstract}




\maketitle

%!TEX root=./main.tex

\section{Introduction}
\seclabel{intro}

Given a nondeterministic finite automaton $\aut$ over an alphabet $\domain$ 
and a length $n$ presented in unary, the counting problem  \cntNFA{}
asks to compute $\norm{\lang{\aut} \cap \domain^n}$, the number of words of length $n$ accepted by $\aut$.  
While the corresponding decision problem (membership in $\lang{\aut}$) 
is solvable in polynomial time, exact counting is $\sharpeP$-hard~\cite{kannan1995counting},
i.e., as hard as $\#\mathsf{SAT}$, the problem of counting the number of 
satisfying assignments to a propositional logic formula.
For counting problems that exhibit computational hardness like this one, 
a natural question to ask is: can we come up with (possibly randomized) 
approximation algorithms that are tractable and offer guarantees on the quality of approximation?
For $\#\mathsf{SAT}$, such an approximation is unlikely (unless $\mathsf{NP}=\mathsf{RP}$~\cite{Dyer2000NPRP}).
For the case of \cntNFA{}, however, the problem has remained open up until recently:
For over two decades the best known approximation algorithm was the quasi-polynomial 
scheme of Kannan, Sweedyk, and Mahaney~\cite{kannan1995counting}. 
But in 2019, Arenas, Croquevielle, Jayaram, and Riveros~\cite{arenas2019efficient,arenas2021nfa} 
(henceforth ACJR) established that \cntNFA{} admits a
fully polynomial-time randomized approximation scheme (FPRAS), 
with running time $\widetilde{O}(n^{17}m^{17}\epsl^{-14}\log(1/\dl))$ for an NFA with $m$ states.  
Subsequent work has substantially improved the complexity: 
Meel, Chakraborty, and Mathur~\cite{meel2024faster} obtained 
$\widetilde{O}\big((n^{10}m^2 + n^6m^3)\epsl^{-4}\log^2(1/\dl)\big)$, 
and de Colnet and Meel~\cite{decolnet2025practical} (henceforth DM) 
obtained $O\big(n^2m^3\log(nm)\,\epsl^{-2}\log(1/\dl)\big)$ via an algorithm based on dependent samples.

The bounds above concern explicitly presented alphabets.
That is, every transition is labeled by one concrete letter from the alphabet.
The running times above hide a polynomial dependence on $\norm{\domain}$.  
This assumption, though, is at odds with how automata arise in applications.  
A transition of a string analyzer, for example, is typically
labeled by a predicate such as ``is a digit'' or ``is not a delimiter'',
and rarely labeled by a single character.  
Similarly, an automaton processing integer inputs may have a
 transition guarded by a condition such as ``the input lies between 0 and 1000''.
Symbolic automata~\cite{veanes2012symbolic,dantoni2017power} naturally capture the 
succinctness desired in these settings by labeling transitions with predicates over $\domain$, 
so that a small number of transitions suffice even when $\domain$ is large.
A predicate labeling a transition is called its \emph{guard}.
The focus of this paper is their counting problem, \cntSNFA{}:

\begin{restatable}{problem}{SymbolicNFACounting}[\cntSNFA{}]
\problabel{symbolic-nfa-counting}
Given a symbolic NFA $\saut$ over alphabet $\domain$ and a natural number $n$ presented in unary, compute $\numwords{\saut}{n} = \norm{\lang{\saut} \cap \domain^n}$.
\end{restatable}
% We remark that a fundamental challenge in \cntSNFA{}, as in \cntNFA{}, is
% that a word may have multiple accepting runs, while the problem asks us to
% count each accepted word exactly once.

For decision problems like membership and emptiness, as well as operations such as intersection and union, 
classical (explicit) automata constructions lift to the symbolic setting, assuming an effective background theory of predicates.
On the other hand, the counting problem for symbolic automata is more delicate.
The difficulty stems from the fact that
each guard may denote many letters, distinct guards may overlap,
and finally, a concrete word may be accepted through many symbolic runs.
Thus, simply counting runs or guard-labelled paths does not suffice.
% If letters have efficiently recognizable, unique encodings of polynomial length
% and guard membership is decidable in polynomial time, then \cntSNFA{} is in
% $\sharpeP$. It is $\sharpeP$-hard for predicate theories that encode ordinary
% NFAs with polynomial overhead.
Indeed, this problem is $\sharpeP$-hard even when the predicates on guards correspond to singleton
sets $\set{0}$ and $\set{1}$.

In another closely related work, Arenas, Croquevielle, Jayaram, and Riveros
studied the problem of counting words of a given length accepted by
\emph{succinct} NFAs. They established an FPRAS assuming membership tests,
approximate counts, and almost-uniform samples for transition
labels~\cite[Theorem~1.4]{arenas2021tree}. This raises the question we study
in this work: \emph{can we extend the recent dependent sampling approach
of DM to symbolic NFAs, obtain substantially improved sampling-query
bounds, and recover DM's running-time bound on constant-size explicit
alphabets?}

%!TEX root=./main.tex

\subsection{Symbolic automata counting in practice}

The \cntSNFA{} naturally arises in a wide range of applications.
In quantitative information flow, one is often interested in computing
the size of the set of inputs corresponding to a given output, and in many applications, this set
is naturally a regular language~\cite{aydin2015automata,BangAPB16}.
In graph databases, the problem of counting paths of a given length that
match regular path queries reduces to the problem of counting accepted
words in finite state automata~\cite{arenas2021nfa}.
Likewise, many testing techniques ask to sample
inputs from a structured set (often a regular language)
according to a uniform or almost-uniform distribution, ensuring an unbiased randomized exploration~\cite{veanes2010rex,gaudel2011counting}.
In all of these domains, the natural alphabets are large or structured, and can be
naturally presented using a few predicates on elements of a fixed domain.

%!TEX root=./main.tex

\begin{figure*}[t]
\centering
\small
% Action colors connect program branches and guards; literals use \strlit.
\definecolor{sanInk}{HTML}{29394D}
\definecolor{sanCopy}{HTML}{17665C}
\definecolor{sanMask}{HTML}{245A9B}
\definecolor{sanEscape}{HTML}{A45414}
\definecolor{sanPaper}{HTML}{F5F7FA}
\definecolor{sanRule}{HTML}{D6DEE7}
\begin{minipage}[t]{0.36\textwidth}
\vspace{0pt}
{\sffamily\bfseries\color{sanInk}(a) Sanitizer $T$}
\begin{lstlisting}[
  basicstyle=\ttfamily\footnotesize\color{sanInk},
  keywordstyle=\bfseries\color{sanInk},
  stringstyle=\bfseries\color{literalInk},
  columns=fullflexible, keepspaces=true, showstringspaces=false,
  frame=single, framerule=0.4pt, frameround=tttt,
  rulecolor=\color{sanRule}, backgroundcolor=\color{sanPaper},
  framesep=4pt, xleftmargin=4pt, xrightmargin=4pt,
  aboveskip=5pt, belowskip=0pt,
  language=Python, escapeinside={(*@}{@*)}]
def sanitize(s):
    out = []
    for c in s:
        if (*@\textcolor{sanMask}{\strlit{"0"} <= c <= \strlit{"9"}}@*):
            (*@\textcolor{sanMask}{out.append(\strlit{"*"})}@*)
        elif (*@\textcolor{sanEscape}{c == \strlit{"<"}}@*):
            (*@\textcolor{sanEscape}{out.append(\strlit{"\&lt;"})}@*)
        else:
            (*@\textcolor{sanCopy}{out.append(c)}@*)
    return "".join(out)
\end{lstlisting}
\end{minipage}
\hfill
\begin{minipage}[t]{0.60\textwidth}
\vspace{0pt}
\centering
{\sffamily\bfseries\color{sanInk}(b) Preimage automaton $\saut_y$}\par
\vspace{5pt}
{\setlength{\fboxsep}{4pt}%
\colorbox{sanPaper}{\makebox[\dimexpr\linewidth-8pt\relax]{%
\footnotesize Observed log\quad
$y=\strlit{ID**\&lt;Z}$}}}\par
\vspace{6pt}
\begin{tikzpicture}[
  >={Stealth[length=1.8mm,width=1.3mm]},
  line cap=round, line join=round,
  state/.style={circle, draw=sanInk!70, fill=sanPaper,
    text=sanInk, line width=0.65pt, minimum size=7mm,
    inner sep=0pt, font=\small},
  accept/.style={state, draw=sanInk, fill=sanInk, text=white,
    double=white, double distance=1.2pt},
  lab/.style={font=\footnotesize, inner sep=2pt},
  copy/.style={->, draw=sanCopy, text=sanCopy, line width=0.85pt},
  mask/.style={->, draw=sanMask, text=sanMask, line width=0.85pt},
  escape/.style={->, draw=sanEscape, text=sanEscape, line width=0.85pt}
]
\node[state]  (r0) at (0,0) {$r_0$};
\node[state]  (r1) at (1.8,0) {$r_1$};
\node[state]  (r2) at (3.6,0) {$r_2$};
\node[state]  (r3) at (5.8,0) {$r_3$};
\node[state]  (r4) at (5.8,-1.5) {$r_4$};
\node[state]  (r5) at (2.9,-1.5) {$r_5$};
\node[accept] (r6) at (0,-1.5) {$r_6$};

\draw[->, draw=sanInk!70, line width=0.85pt] (-0.65,0) -- (r0);
\draw[copy] (r0) -- node[above, lab] {$c=\strlit{I}$} (r1);
\draw[copy] (r1) -- node[above, lab] {$c=\strlit{D}$} (r2);
\draw[mask] (r2) -- node[above, lab] {$\mathsf{Digit}(c)$} (r3);
\draw[mask] (r3) -- node[right, lab] {$\mathsf{Digit}(c)$} (r4);
\draw[escape] (r4) -- node[above, lab] {$c=\strlit{<}$} (r5);
\draw[copy] (r5) -- node[above, lab] {$c=\strlit{Z}$} (r6);
\end{tikzpicture}
\par\vspace{3pt}
{\footnotesize
$\textcolor{sanMask}{\mathsf{Digit}(c)}
\iff c\in\{\strlit{0},\ldots,\strlit{9}\}$.\par
\vspace{3pt}
Input length $n=6$\qquad
$\textcolor{sanMask}{10\times10=100}$ compatible secrets.}
\end{minipage}

\caption{A sanitizer and the secrets consistent with the observed log
$y=\strlit{ID**\&lt;Z}$. Literal characters and strings use plum monospace.
Teal, blue, and orange link copying, digit masking, and escaping across
the program and automaton. The two digit guards yield $100$ possible secrets.}
\Description{Two panels show a sanitizer program and a seven-state
symbolic automaton. The program copies ordinary characters, replaces
ASCII digits with asterisks, and escapes a less-than sign as an HTML
entity. For the observed log ID followed by two asterisks, the escaped
less-than sign, and Z, the automaton accepts ID followed by two ASCII
digits, a less-than sign, and Z. It accepts exactly 100 words of length
six. Teal, blue, and orange connect the copying, masking, and escaping
operations across the panels. Literal characters and strings use bold
plum monospace.}
\figlabel{sanitizer-preimage}
\end{figure*}
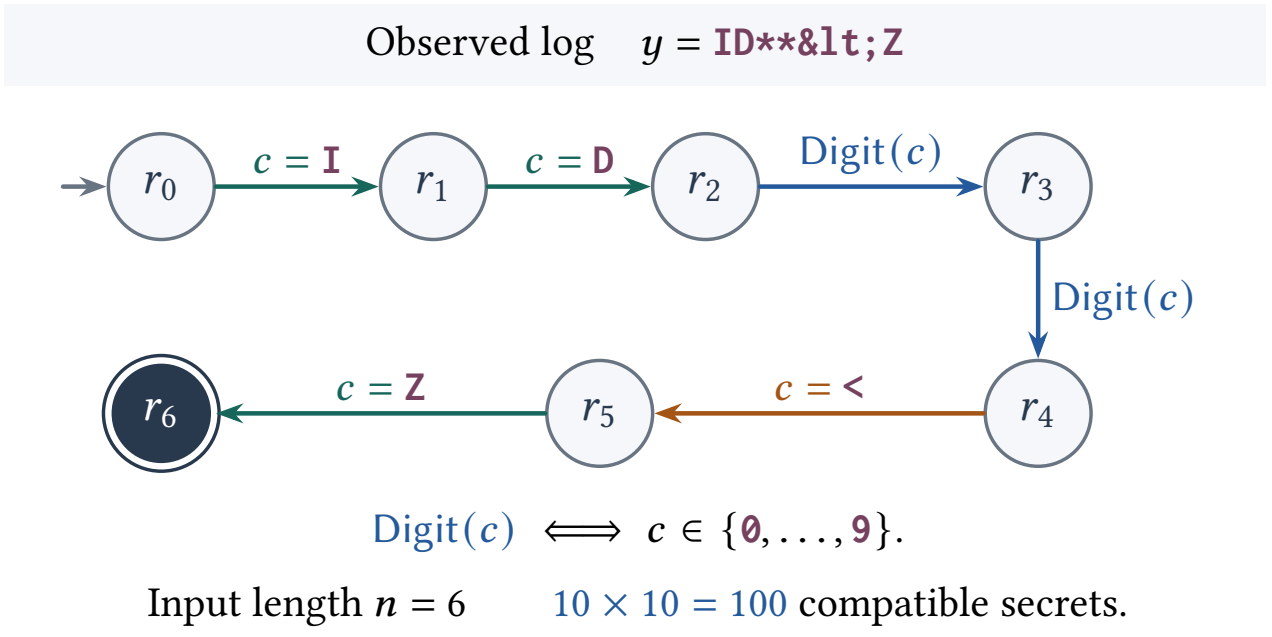

\figref{sanitizer-preimage}(a) illustrates a standard
sanitizer-analysis pattern~\cite{HooimeijerLMSV11,Veanes13}.
A web service stores a secret request identifier \texttt{s} but writes only
its sanitized value to a low-privilege log. We model the sanitizer as a
\emph{symbolic} finite-state transducer $T$ that copies ordinary identifier
characters, replaces each ASCII digit with the public mask symbol
\strlit{*}, and escapes each \strlit{<} as the four-character string
\strlit{\&lt;}.
To keep the presentation simple, we assume that \texttt{s} contains no
literal \strlit{*} and that every occurrence of \strlit{\&lt;} in the
output results from escaping an input \strlit{<}.

Once an attacker observes a public log string $y$, the possible values of
the secret identifier form the preimage language
$L_y = \setpred{s \in \domain^*}{T(s)=y}$. For the observation
$y=\strlit{ID**\&lt;Z}$, \figref{sanitizer-preimage}(b) shows a symbolic
automaton $\saut_y$ recognizing $L_y$, with edge colors matching the program
branches. The copied characters \strlit{ID} and \strlit{Z}
are fixed, each mask \strlit{*} allows any of the ten ASCII digits, and
the escape \strlit{\&lt;} corresponds to a single input \strlit{<}.
The two digit positions can be chosen independently, giving
$10\cdot 10=100$ possible inputs, each of length $n=6$.
In general, counting the secrets of length $n$ consistent with an observation reduces to
symbolic-automata model counting~\cite{aydin2015automata}: computing
$\norm{\lang{\saut_y}\cap \domain^n}$. Comparing this count with the prior number of possible secrets
helps quantify information leakage~\cite{BangAPB16}.

%!TEX root=./main.tex

\subsection{Overview of our contributions}

There are two natural ways to reduce \cntSNFA{} to counting over an explicit
alphabet. The first is to expand every guard into the concrete letters
satisfying it. This introduces a dependence on $\norm{\domain}$, which may be
very large (e.g., $2^{64}$ for a domain of 64-bit integers). The second is to
partition the alphabet into \emph{minterms} (Boolean atoms) and construct a
weighted explicit automaton.
While this can be done when the predicate theory supports
constructing and counting Boolean combinations of guards, the number
of minterms can still be exponential in the number of guards. Thus neither reduction
guarantees a running time polynomial in the size of the symbolic description.

We avoid these reductions by adapting the dependent-sampling approach of DM
directly to symbolic automata. To state our result, we first specify what access
the algorithm requires to the predicate theory. We use a weak interface, which we call \emph{$\dtol$-quantitative access} (\defref{quantitative-access}): for each guard $\predphi$ appearing on a transition, the algorithm may (i) test membership $a \sat \predphi$ of  concrete letters $a \in \domain$, (ii) obtain an \emph{approximate} count of $\denote{\predphi}$ within a multiplicative factor $1 \pm \dtol$, and (iii) draw \emph{almost}-uniform samples from $\denote{\predphi}$, with each letter's sampling probability within a multiplicative factor $1\pm\dtol$ of its uniform probability.  We note that the interface is required only for the guards that actually appear in the input automaton, i.e., our algorithm does not construct more predicates (for example, arbitrary minterms).  Our result also permits infinite ambient alphabets, provided every transition guard has finite denotation.  The key contribution of this work is to establish the following theorem (stated formally as \thmref{symbolic-fpras}).

\myparagraph{Main result (informally)}
There is a randomized algorithm that, given a symbolic NFA $\saut$ with state set $\states$ and transition set $\trans$, a length $n$ in unary, rational $\epsl \in (0,1]$ and $\dl \in (0,1)$, and $\dtol$-quantitative access to the guards of $\saut$ with $\dtol \leq \epsl/(400(n+2)^2)$, returns a nonnegative integer estimate $\widetilde{N}$ with
\[
    \Pr\big[(1-\epsl)\,\numwords{\saut}{n} \leq \widetilde{N} \leq (1+\epsl)\,\numwords{\saut}{n}\big] \geq 1-\dl,
\]
using $O(\norm{\trans})$ counting queries,
$O\big(n^2\norm{\trans}\allowbreak\log(n\norm{\states})\,\allowbreak\epsl^{-2}\log(1/\dl)\big)$
sampling queries, and
$O\big(n^2\norm{\trans}\,\allowbreak(\dmax +
    \norm{\states})\allowbreak\log(n\norm{\states})\,\allowbreak\epsl^{-2}\log(1/\dl)\big)$
membership queries; here $\dmax$ is the maximum in-degree of a state.
Observe that, with unit-cost oracle operations, the running time has the same
bound as the membership-query count.

We measure running times in the word-RAM model with access to independent
uniform random words, storing each letter in one machine word. In
\secref{approx}, we account for the cost of arithmetic on intermediate estimates
that occupy multiple machine words.

For an explicit alphabet of constant size,
many of the above quantities are bounded as: $\norm{\trans} \in O(\norm{\states}^2)$ and
$\dmax \in O(\norm{\states})$, and thus the running time is
$O\big(n^2\norm{\states}^3\log(n\norm{\states})\,
\epsl^{-2}\log(1/\dl)\big)$, matching the best known bound for
\cntNFA{}~\cite{decolnet2025practical}.
More importantly, for finite alphabets, whenever the predicate theory supports the three operations in time
$\poly{\log\norm{\domain}, 1/\dtol}$ (as is the case for intervals, character
classes, bit-vector ranges, and finite products thereof),
\probref{symbolic-nfa-counting} admits an FPRAS.

The standard interface for symbolic automata~\cite{dantoni2017power,veanes2021automata}
provides Boolean operations on guards and satisfiability tests. These suffice
for emptiness, product, determinization, and minimization. Counting, however,
requires information about how many letters satisfy each guard.
One could require exact counts and uniform samples, but exact guard counting
is $\sharpeP$-hard for expressive theories such as bit-vectors, where it subsumes
$\#\mathsf{SAT}$~\cite{chistikov2015approximate,chakraborty2016approx}.
We therefore require only approximate counts and almost-uniform samples within
a tolerance $\dtol$
(\defref{quantitative-access}).

For interval and character-class guards~\cite{brics-automaton,saarikivi2019symbolic,moseley2023derivative},
counting and sampling can be performed exactly. When a guard is represented as
a union of disjoint ranges, its count is the sum of the range sizes. To sample
uniformly, we choose a range with probability proportional to its size and then
a letter uniformly from that range. For more expressive guards, such as Boolean
and bit-vector formulas, we can use hashing-based approximate counters and
almost-uniform samplers~\cite{chistikov2015approximate,chakraborty2016approx,delannoy2022almost}.

Of course, our theoretical algorithm will need some more work
to translate to a practical algorithm.
For small alphabets, explicit automata may still be simpler to
implement and avoid the overhead of predicate operations. Mintermization or
bit-blasting can also be effective when the resulting explicit representation
is small. The \textsc{FastNFA} implementation~\cite{heissl2026fastnfa}
demonstrates that an FPRAS for explicit NFAs can be effective in practice. The
performance of a symbolic implementation of our algorithm would also depend on
the cost of counting and sampling from guards.

\myparagraph{Mechanization}
We have also formalized the core probabilistic correctness argument in Lean.
It is publicly available~\cite{mathur2026symbolicartifact}. 
We used AI assistance, guided by
our pen-and-paper mathematical proofs. We reviewed the Lean theorem
statements and proof structure.

To summarize, our contributions are:
\begin{enumerate}[leftmargin=*]
    \item \emph{Problem formulation.} We formulate \cntSNFA{} under an explicit quantitative interface to the predicate theory (\secref{prelim}, \secref{interface}).
    \item \emph{Algorithm.} We extend DM by sampling one letter per stored word and transition, incorporating guard counts into retention probabilities, indexing resampling caps by transitions, and resolving overlaps through membership-based canonical-transition tests (\secref{approx}).
    \item \emph{Analysis.} We bound joint moments by conditioning on the sampling history, accounting for collisions and multiplicative oracle bias (\secref{analysis}).
    \item \emph{Machine-checked correctness.} We formalize the core probabilistic correctness chain in the Lean 4 proof assistant.
\end{enumerate}

The rest of the paper is organized as follows. \secref{prelim} introduces
symbolic NFAs. \secref{overview} conveys the main ideas of our approach.
\secref{approx} presents the quantitative interface, the algorithm, and its
cost accounting, and \secref{analysis} carries out the analysis.
\secref{related} discusses related work. We conclude in \secref{conclusion}.

%!TEX root=./main.tex

\section{Preliminaries}
\seclabel{prelim}

In this section, we fix notation for words and symbolic automata, state the
counting problem, and record the probabilistic facts used in \secref{analysis}.

\myparagraph{Words and slices}{
    Let $\domain$ be a set, called the concrete alphabet.  A word over $\domain$ is a finite sequence $w = a_1a_2\ldots a_k$ of letters $a_i \in \domain$; its length is $\len{w} = k$, and the empty word $\emptyword$ has length $0$.  We write $\domain^*$ for the set of all words, $\domain^n$ for the words of length $n$, $w \concat a$ for the word $w$ extended by the letter $a$, and $w^j$ for the length-$j$ prefix of $w$.  
    % For a language $L \subseteq \domain^*$, the $n^{\text{th}}$ \emph{slice} of $L$ is $L_n = L \cap \domain^n$.
}

\myparagraph{Symbolic NFAs}{
    A symbolic alphabet is a tuple $\calD = (\domain, \predset, \denote{\cdot})$, where $\predset$ is a set of predicates and each $\predphi \in \predset$ denotes a set $\denote{\predphi} \subseteq \domain$; we write $a \sat \predphi$ when $a \in \denote{\predphi}$, and $\cnt{\predphi} = \norm{\denote{\predphi}}$. 
    A symbolic nondeterministic finite automaton (symbolic NFA) over $\calD$ is a tuple $\saut = (\states, \predset, \trans, \qI, \accept)$, where $\states$ is a finite set of states, $\qI \in \states$ is the initial state, $\accept \subseteq \states$ is the set of accepting states, and $\trans \subseteq \states \times \predset \times \states$ is a finite transition relation. 
     A run of a word $w = a_1a_2\ldots a_n \in \domain^n$ on $\saut$ is a sequence $q_0,q_1,\ldots,q_n$ such that $q_0 = \qI$ and, for every $1 \leq i \leq n$, there is a transition $(q_{i-1},\predphi_i,q_i) \in \trans$ with $a_i \sat \predphi_i$; the run is accepting if $q_n \in \accept$.  The language $\lang{\saut} \subseteq \domain^*$ is the set of concrete words that have an accepting run on $\saut$. 
    A predicate labeling a transition is called its \emph{guard}.  We assume that the denotation $\denote{\predphi}$ of every predicate $\predphi$ occurring as a guard in the input automaton is finite.  The algorithm of \secref{approx} uses membership tests, \emph{approximate} counting, and \emph{almost}-uniform sampling only over these guards (\defref{quantitative-access}).
      An ordinary NFA is the special case in which every guard denotes a singleton, and \cntNFA{} is the corresponding counting problem.
}

\SymbolicNFACounting*

Since the automaton has finitely many transitions and each guard has finite
denotation, the union of its guard denotations is finite, and hence so is
every fixed-length slice of its language.
\cntSNFA{} is $\sharpeP$-hard under polynomial-time Turing reductions,
even when every guard denotes $\set{\strlit{0}}$ or $\set{\strlit{1}}$.
This follows from the $\sharpeP$-hardness of binary
\cntNFA{}~\cite{arenas2021nfa}: replacing each transition label by its
singleton guard preserves the accepted language and the requested word
length.

\myparagraph{FPRAS and almost-uniform sampling}{
    A fully polynomial-time randomized approximation scheme (FPRAS) for \probref{symbolic-nfa-counting} is a randomized algorithm that, given $\saut$, unary $n$, and $\epsl \in (0,1]$, $\dl \in (0,1)$, runs in time polynomial in the input size, $n$, $\epsl^{-1}$, and $\log(\dl^{-1})$, and returns a nonnegative integer estimate $\widetilde{N}$ with $\Pr\big[(1-\epsl)\numwords{\saut}{n} \leq \widetilde{N} \leq (1+\epsl)\numwords{\saut}{n}\big] \geq 1-\dl$, where the probability is over the algorithm's random choices.  The corresponding almost-uniform sampling problem asks for a randomized algorithm that samples words from a distribution within total variation distance $\dl$ of the uniform distribution over $\lang{\saut} \cap \domain^n$.
}

\myparagraph{Unrolled symbolic automaton}{
    Our algorithm operates on the unrolling of the input automaton.  For $n \in \nats$, the unrolled automaton $\sautunroll_n$ has states $q^i$ for $q \in \states$ and $0 \leq i \leq n$ (we say $q^i$ lies in \emph{layer} $i$), initial state $\qI^0$, accepting states $\setpred{q^n}{q \in \accept}$, and transitions $(q^i,\predphi,(q')^{i+1})$ for $(q,\predphi,q') \in \trans$ and $0 \leq i < n$; it has $(n+1)\norm{\states}$ states and $n\norm{\trans}$ transitions and is constructed in time $O(n\norm{\trans})$.  We write $\langstate{q^i}$ for the set of words of length $i$ that can take $\sautunroll_n$ from $\qI^0$ to $q^i$, and $\intrans{q^i}$ for the set of incoming transitions of $q^i$.  For a transition $t = ((q')^{i-1},\predphi,q^i)$, we use $\langstate{t} = \setpred{w\concat a}{w \in \langstate{(q')^{i-1}},\ a \in \denote{\predphi}}$ to denote the sublanguage of $\langstate{q^i}$ of words that have a run to $q^i$ through $t$; thus $\langstate{q^i} = \bigcup_{t \in \intrans{q^i}} \langstate{t}$.
}

\myparagraph{Probability background}{
    All random variables in this paper are defined on the discrete probability space generated by the algorithm's random choices (oracle draws and coins).  For a collection $\calG$ of random variables, $\Ex[X \mid \calG]$ denotes the conditional expectation of $X$ given the values of all variables in $\calG$; it is itself a random variable, namely a function of $\calG$, and $\Ex[X \mid \emptyset] = \Ex[X]$.  We say that $X$ is \emph{fixed} by $\calG$ when $X$ is a function of the variables in $\calG$.  We freely use three standard facts~\cite{mitzenmacher2017probability}:
    \begin{enumerate}[leftmargin=*]
        \item[(a)] the \emph{tower rule}: $\Ex\big[\Ex[X \mid \calH] \,\big|\, \calG\big] = \Ex[X \mid \calG]$ whenever $\calG \subseteq \calH$;
        \item[(b)] the \emph{fixed-variable rule}: $\Ex[XY \mid \calG] = X\,\Ex[Y \mid \calG]$ whenever $X$ is fixed by $\calG$;
        \item[(c)] \emph{conditional independence}: if $X_1,\ldots,X_k$ are functions of pairwise disjoint collections of random variables, each independent of the others and of $\calG$ (possibly together with variables fixed by $\calG$), then $\Ex\big[\prod_{i=1}^k X_i \,\big|\, \calG\big] = \prod_{i=1}^k \Ex[X_i \mid \calG]$.
    \end{enumerate}
}

Our analysis also uses two concentration inequalities:

\begin{proposition}[{\cite{mitzenmacher2017probability}}]
    \proplabel{concentration-tools}
    (a)~(Markov) If $X \geq 0$ and $a > 0$, then $\Pr[X \geq a] \leq \Ex[X]/a$.
    (b)~(Hoeffding) If $X_1,\ldots,X_\mu$ are independent random variables with values in $[0,1]$ and $\bar X = X_1 + \cdots + X_\mu$, then $\Pr\big[\bar X - \Ex[\bar X] \geq t\big] \leq e^{-2t^2/\mu}$ for every $t \geq 0$.
\end{proposition}

%!TEX root=./main.tex

\section{Overview of the Approach}
\seclabel{overview}

We explain how to adapt the dependent-sampling algorithm of DM to symbolic guards. The main changes are sampled extensions, a canonical-transition rule for overlapping guards, and an analysis that accounts for collisions and oracle error.

\myparagraph{Representing languages by small sample sets}{
Assume for now that guard counts are exact and guard sampling is uniform.
In the unrolled automaton, a transition $t=(q',\predphi,q)$ contributes
$\langstate{t}=\setpred{v\concat a}{v\in\langstate{q'},\ a\sat\predphi}$.
For every noninitial state $q$,
\[
\langstate{q}=\bigcup_{t\in\intrans{q}}\langstate{t},
\qquad
\norm{\langstate{t}}=\norm{\langstate{q'}}\cnt{\predphi}.
\]
These transition languages may overlap, so summing their sizes can
overcount, while storing their union can require exponential space.

Following DM~\cite{decolnet2025practical}, we represent these languages by
small random subsets. If each word of a finite language $L$ belongs to a
random subset $S$ with fixed probability $p>0$, then
$\Ex[\norm{S}] = p\norm{L}$, so $\norm{S}/p$ is an unbiased estimate of
$\norm{L}$. Choosing $p$ close to $1/\norm{L}$ keeps the expected sample
size constant.

At each state $q$, we estimate $1/\norm{\langstate{q}}$ by $p(q)$ and
maintain sample sets
$S^1(q),\ldots,S^\alpha(q) \subseteq \langstate{q}$. We construct them
layer by layer, extending stored prefixes from predecessor states and
estimating language sizes from the resulting samples. Reusing prefixes
makes the sample sets and estimates share random choices. Consequently,
concentration bounds for independent samples do not directly establish the
accuracy of these estimates. DM's dependent sampling approach gains
efficiency from this reuse and controls the resulting dependence in its
analysis.
Symbolic guards additionally require extending
prefixes without enumerating the letters satisfying a guard.
}

\myparagraph{Extending samples without enumerating guards}{
To combine contributions from different incoming transitions, we aim for
a common sampling scale $\rho(q)$. For a transition $t=(q',\predphi,q)$, we
could enumerate every extension of a stored prefix $v\in S^r(q')$ and retain each
with probability $\rho(q)/p(q')$. Instead, we draw one uniformly random
letter $a\in\denote{\predphi}$ and retain $u=v\concat a$ in
$\Sbar^r(t)$ with probability
$\rho(q)\cnt{\predphi}/p(q')$. Conditional on the preceding sampling
history, each particular extension of $v$ is therefore generated and
retained with probability
\[
\frac{1}{\cnt{\predphi}}
\cdot\frac{\rho(q)\cnt{\predphi}}{p(q')}
=\frac{\rho(q)}{p(q')}.
\]
Thus each extension has the same individual retention probability as
under enumeration, while we generate only one candidate per stored prefix
and transition. Choosing
$\rho(q)=\min_{(q',\predphi,q)\in\intrans{q}}p(q')/\cnt{\predphi}$
ensures that all retention probabilities are at most one. We use fresh
randomness for every triple $(r,t,v)$. It remains to
combine these candidates without overrepresenting words that can arrive
through several transitions.
}

{
\begin{wrapfigure}{R}{0.32\textwidth}
\centering
% Match the state styling and palette of Figure 1.
\definecolor{unionInk}{HTML}{29394D}
\definecolor{unionFirst}{HTML}{17665C}
\definecolor{unionSecond}{HTML}{245A9B}
\definecolor{unionPaper}{HTML}{F5F7FA}
\begin{tikzpicture}[
  >={Stealth[length=1.8mm,width=1.3mm]},
  line cap=round, line join=round,
  state/.style={circle, draw=unionInk!70, fill=unionPaper,
    text=unionInk, line width=0.65pt, minimum size=7mm,
    inner sep=0pt, font=\small},
  target/.style={state, draw=unionInk, fill=unionInk, text=white},
  lab/.style={font=\small, fill=white, inner sep=2pt},
  first/.style={->, draw=unionFirst, text=unionFirst, line width=0.85pt},
  second/.style={->, draw=unionSecond, text=unionSecond, line width=0.85pt}
]
\node[state] (q1) at (0,0.65) {$q_1$};
\node[state] (p1) at (0,-0.65) {$q'_1$};
\node[target] (q2) at (2.55,0) {$q_2$};
\draw[first] (q1) -- node[pos=0.45,above=2pt,lab] {$\predphi$} (q2);
\draw[second] (p1) -- node[pos=0.45,below=2pt,lab] {$\predphi'$} (q2);
\end{tikzpicture}
\caption{Two guarded transitions into $q_2$.}
\label{figure:automaton}
\Description{Two lightly shaded predecessor states point to a dark
successor state q2. The upper teal transition is labeled phi and the lower
blue transition is labeled phi prime. All three states are ordinary
single-circle states.}
\end{wrapfigure}
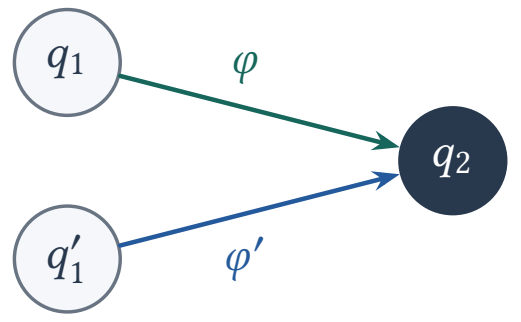

\myparagraph{Canonical union}{
Merging the transition sample sets can give words with several incoming
routes extra chances of inclusion. Following DM, we fix an order on
incoming transitions and retain a candidate only when it arrives through
the first transition that could produce it---its \emph{canonical
transition}. In Figure~\ref{figure:automaton}, suppose the upper transition
precedes the lower one. A candidate $v\concat a$ generated through the
lower transition is discarded whenever $v\in\langstate{q_1}$ and
$a\sat\predphi$, even if the upper transition did not generate it. Each
word can therefore enter the sample only through its canonical transition.
To implement this test, we cache the states reachable by each stored
prefix. Checking an alternative transition then requires a cache lookup
and a guard-membership query, without constructing Boolean combinations
of guards.
}

\par
}

\myparagraph{Estimating counts and preparing the next layer}{
Let $\Shat^r(q)$ denote the sets obtained after canonical union.
The normalized sizes $\norm{\Shat^r(q)}/\rho(q)$ provide estimates of
$\norm{\langstate{q}}$. We divide the $\alpha=\beta\gamma$ estimates
into $\gamma$ blocks of $\beta$ sets, and write $Y_{q,b}$ for the average
in block $b$. The reciprocal median $\hat\rho(q)$ gives the next sampling scale:
\[
\hat\rho(q)=\median(Y_{q,1},\ldots,Y_{q,\gamma})^{-1},
\qquad p(q)=\minsf\{\rho(q),\hat\rho(q)\}.
\]
We then
retain each word of each $\Shat^r(q)$ independently with probability
$p(q)/\rho(q)$, obtaining the sets $S^r(q)$ used by successor states.
The cap ensures that this retention probability is at most one, while
thinning controls the size of the sets passed to the next layer.
}

\myparagraph{Controlling dependence}{
To establish the accuracy of the median, we must control the probability
that many blocks are inaccurate simultaneously. Following
DM~\cite{decolnet2025practical}, we bound joint moments of their errors.
The key idea is to trace two words back to their last common ancestor:
a prefix at a state shared by both canonical runs. Within one repetition,
their contributions share the sampling decisions along this common
ancestry; we bound their correlation by analyzing how the runs separate.
Formally, for distinct
$w_1,w_2\in\langstate{q}$, their \emph{divergence level}
$k=\Div(w_1,w_2)$ is the last layer where both prefixes and states agree;
the common state
$\divv{w_1}{w_2}$ is their \emph{divergence state}. At layer $k+1$, distinct
transitions use fresh draws that are independent conditional on the
preceding history. If instead the same transition requires two different
letters, the words compete for a single draw and cannot both enter the
same sample set $S^r(q)$. We call this a \emph{collision}; such pairs have zero
joint inclusion probability. For the remaining pairs, the bound depends
on $\norm{\langstate{\divv{w_1}{w_2}}}$. A counting argument bounds the
total contribution at each divergence level $k$; summing over levels
introduces only a factor linear in $n$. Thus the relative variance of
$Y_{q,b}$ lies in $O(n/\beta)$. Taking $\beta\in\Theta(n/\epsl^2)$ gives
constant-probability accuracy. The joint-moment bounds then establish
concentration of the median despite dependence between blocks.
}

\myparagraph{Handling approximate oracles}{
We now use a guard-count estimate $\acnt{\predphi}$ with relative error
$\dtol$, and a sampler whose letter probabilities $\sdist{\predphi}(a)$
lie within a factor $1\pm\dtol$ of $1/\cnt{\predphi}$. Each extension
now carries the factor $\sdist{\predphi}(a)\acnt{\predphi}$, equal to $1$
with exact oracles. Multiplying these factors along the canonical
run gives a word weight $\wgt(w,q)$, and our analysis tracks the weighted
mass $\mass{q}=\sum_{w\in\langstate{q}}\wgt(w,q)$. We carry these weights
through the variance and joint-moment bounds, while adjusting the accuracy
targets across layers to accommodate the accumulated error. An oracle
tolerance $\dtol\in O(\epsl/n^2)$ suffices to preserve the desired
relative-error guarantee.
}

%!TEX root=./main.tex

\section{An FPRAS for Symbolic NFAs}
\seclabel{approx}

We now present the algorithm and its query bounds. We first specify the guard interface, then describe preprocessing, state processing, and the cache. Correctness is proved in \secref{analysis}.

\subsection{The Quantitative Interface}
\seclabel{interface}

\begin{definition}[$\dtol$-quantitative access]
\deflabel{quantitative-access}
Let $\saut$ be a symbolic NFA over the symbolic alphabet $\calD$ and let $\dtol \in [0,1)$.  We say that the algorithm has \emph{$\dtol$-quantitative access} to $\saut$ when, for every guard $\predphi$ appearing in $\trans$, the following three operations are available.
\begin{enumerate}[leftmargin=*]
\item[(i)] \emph{Membership} $\omember(a,\predphi)$: given $a \in \domain$, decide whether $a \sat \predphi$, in time $\Tmem$.
\item[(ii)] \emph{Approximate counting}: a value $\acnt{\predphi} \in \nats$ satisfying
\[
(1-\dtol)\cnt{\predphi} \;\leq\; \acnt{\predphi} \;\leq\; (1+\dtol)\cnt{\predphi},
\]
computable in time $\Tcnt$.
\item[(iii)] \emph{Almost-uniform sampling} $\osample{\predphi}$: when $\cnt{\predphi} > 0$, a randomized procedure running in time $\Tsamp$ that returns a letter of $\denote{\predphi}$; successive calls are mutually independent and follow a fixed distribution $\sdist{\predphi}$ over $\denote{\predphi}$ satisfying, for every $a \in \denote{\predphi}$,
\[
\frac{1-\dtol}{\cnt{\predphi}} \;\leq\; \sdist{\predphi}(a) \;\leq\; \frac{1+\dtol}{\cnt{\predphi}}.
\]
\end{enumerate}
The values $\acnt{\predphi}$ are queried once per guard and memoized, so that the same guard always reports the same count.\footnote{Randomized counters suffice: assign failure probability $\dl/(2\norm{\trans})$ to each memoized count and $\dl/2$ to the algorithm, and apply a union bound.}
\end{definition}

Setting $\dtol = 0$ requires exact counting and uniform sampling.  \thmref{symbolic-fpras} requires $\dtol \leq \epsl/(400(n+2)^2)$.  For finite alphabets, a predicate theory supporting the three operations at cost $\poly{\log\norm{\domain}, 1/\dtol}$ yields an FPRAS.  Guarantee (iii) is multiplicative rather than in total variation distance: this is what standard almost-uniform generators provide, and it is what the analysis of \secref{analysis} needs.
%
%\begin{remark}
%\deflabel{randomized-counter}
%If the counter in (ii) is randomized and meets its guarantee only with probability $1-\dl'$ per call, we invoke it with $\dl' = \dl/(2\norm{\trans})$, union-bound over the $\norm{\trans}$ memoized calls, and run the algorithm with failure probability at most $\dl/2$; we therefore assume the counts are deterministic.
%\end{remark}

We use the word-RAM model~\cite{fredman1993fusion}, augmented with independent
uniform random words~\cite{bringmann2013exact}. Arithmetic on machine words,
including multiplication and integer division, comparisons, shifts, and
random-word generation take constant time. In this model, we assume that
each letter occupies one machine word, and that guard counts, parameters,
and indices occupy a constant number of words. We store larger integers
across multiple words and represent estimates as exact fractions.
In complexity bounds, logarithmic factors are bounded below by $1$; we
handle $n=0$ and empty transition relations directly, and state the bounds
for the remaining cases.

\begin{theorem}[Main]
\thmlabel{symbolic-fpras}\leanchecked
There is a randomized algorithm $\approxalg$ that, given a symbolic NFA $\saut = (\states,\predset,\trans,\qI,\accept)$, a length $n$ in unary, rational $\epsl \in (0,1]$ and $\dl \in (0,1)$, and $\dtol$-quantitative access to $\saut$ with $\dtol \leq \epsl/(400(n+2)^2)$, returns a nonnegative integer estimate $\widetilde{N}$ with $\Pr\big[(1-\epsl)\,\numwords{\saut}{n} \leq \widetilde{N} \leq (1+\epsl)\,\numwords{\saut}{n}\big] \geq 1-\dl$, where the probability is over the algorithm's random choices.  Writing $\dmax$ for the maximum in-degree of a state of $\saut$, the algorithm makes $Q_{\mathsf{cnt}} \in O(\norm{\trans})$ counting queries, $Q_{\mathsf{samp}} \in O\big(n^2\norm{\trans}\log(n\norm{\states})\,\epsl^{-2}\log(1/\dl)\big)$ sampling queries, and $Q_{\mathsf{mem}} \in O\big(n^2\norm{\trans}(\dmax + \norm{\states})\log(n\norm{\states})\,\epsl^{-2}\log(1/\dl)\big)$ membership queries, and runs in time $O\big(Q_{\mathsf{cnt}}\Tcnt + Q_{\mathsf{samp}}\Tsamp + Q_{\mathsf{mem}}(\Tmem + 1)\big)$ on the word RAM described above.
\end{theorem}

\begin{corollary}[Unit-cost oracles]
\corlabel{unit-cost}
If $\Tcnt, \Tsamp, \Tmem \in O(1)$, then $\approxalg$ makes
$O\big(n^2\norm{\trans}\allowbreak\log(n\norm{\states})\,\allowbreak\epsl^{-2}\log(1/\dl)\big)$
counting and sampling queries and runs in time
$O\big(n^2\norm{\trans}(\dmax + \norm{\states})\allowbreak\log(n\norm{\states})\,\allowbreak\epsl^{-2}\log(1/\dl)\big)$.
\end{corollary}

%\begin{remark}
%\deflabel{dcm-comparison}
%We incur the factor $\dmax + \norm{\states}$ between the counting and sampling query bound and the unit-cost running-time bound when performing canonical-transition tests and cache updates. For these operations, we use both membership queries and additional bookkeeping.  Instantiated on an explicit alphabet of constant size, where $\norm{\trans} \in O(\norm{\states}^2)$ and $\dmax \in O(\norm{\states})$, the running time of \corref{unit-cost} is $O(n^2\norm{\states}^3\log(n\norm{\states})\,\epsl^{-2}\log(1/\dl))$, matching the bound of~\cite{decolnet2025practical}.
%\end{remark}

\subsection{Normal Form and Preprocessing}
\seclabel{normalform}

If $n=0$, we return $1$ when $\qI\in\accept$ and $0$ otherwise. For $n>0$, an empty transition relation gives count $0$. We henceforth assume $n\geq1$ and $\trans\neq\emptyset$.

We first arrange for a single final state.  Let $\eos \notin \domain$ be a fresh end-marker letter and extend the concrete alphabet to $\domain \cup \set{\eos}$ together with a predicate $\predphi_\eos$ with $\denote{\predphi_\eos} = \set{\eos}$; its quantitative interface is trivial ($\acnt{\predphi_\eos} = 1$, the sampler returns $\eos$, membership is an equality test) and exact.  We add a fresh state $\qF$ to $\saut$ with transitions $(q,\predphi_\eos,\qF)$ for every $q \in \accept$, and make $\qF$ the unique accepting state.  Every word of $\lang{\saut} \cap \domain^n$ extends uniquely to a word of length $n+1$ ending in $\eos$ that is accepted at $\qF$, hence counting length-$(n+1)$ words of the modified automaton is exactly \probref{symbolic-nfa-counting}.  We rename the modified automaton $\saut$ and count words of length $n+1$; the notation $\numwords{\saut}{n}$ continues to denote the original input count.

Next, the algorithm computes the unrolled automaton $\sautunroll_{n+1}$ of \secref{prelim}, whose states are stratified into layers $\states^0, \states^1, \ldots, \states^{n+1}$ with $\states^0 = \set{\qI}$, and \emph{trims} it: transitions whose guard $\predphi$ has $\acnt{\predphi} = 0$ are removed and states that are not reachable from $\qI$ or not co-reachable from $\qF$ are removed.  Trimming takes time $O(n\norm{\trans})$ after the $\norm{\trans}$ memoized counting calls.  If $\qF$ is trimmed away then $\numwords{\saut}{n} = 0$ and the algorithm returns $0$; we therefore assume from now on that every state $q$ of $\sautunroll_{n+1}$ satisfies $\langstate{q} \neq \emptyset$ and every guard $\predphi$ in $\sautunroll_{n+1}$ satisfies $\cnt{\predphi} \geq 1$.

We write $\statesunroll$ and $\transunroll$ for the states and transitions of the trimmed unrolled automaton, and we fix (arbitrarily) a total processing order $\prec$ on $\statesunroll$ that refines the layer order, as well as a total order $\prec$ on $\intrans{q}$ for every $q$.  Recall from \secref{prelim} that for a transition $t = (q',\predphi,q) \in \transunroll$ we write $\langstate{t} = \setpred{v \concat a}{v \in \langstate{q'},\, a \in \denote{\predphi}}$, so that $\langstate{q} = \bigcup_{t \in \intrans{q}} \langstate{t}$.  Two observations about this decomposition underlie the algorithm.  First, since distinct pairs $(v,a)$ yield distinct words $v \concat a$,
\begin{equation}
\equlabel{transition-product}
\norm{\langstate{t}} = \norm{\langstate{q'}} \cdot \cnt{\predphi}
\qquad\text{for every } t = (q',\predphi,q) \in \transunroll,
\end{equation}
and in particular $\norm{\langstate{q}} \geq \norm{\langstate{q'}}\cdot\cnt{\predphi}$.  Second, the union over $\intrans{q}$ is in general \emph{not} disjoint: guards of distinct incoming transitions may overlap, and the same word may reach $q$ through several predecessors.  The first observation replaces, in the symbolic setting, the trivial fact $\norm{\langstate{q}} \geq \norm{\langstate{q'}}$ used in the explicit-alphabet FPRAS of~\cite{decolnet2025practical}; the second is the reason for the canonical-transition construction below.

\myparagraph{Canonical transitions}{
For $q \in \states^i$ with $i > 0$ and a word $w = v \concat a \in \langstate{q}$, we define the \emph{canonical transition} of $(w,q)$ as
\[
\canon(w,q) = \min\nolimits_{\prec}\, \setpred{(q',\predpsi,q) \in \intrans{q}}{v \in \langstate{q'} \text{ and } a \sat \predpsi},
\]
and the \emph{canonical child} of $(w,q)$ as the pair $(v,q')$ where $\canon(w,q) = (q',\predpsi,q)$.  Iterating the canonical child relation yields the \emph{canonical run} $\cpath{w}{q} = (q^0_w, q^1_w, \ldots, q^i_w)$ with $q^0_w = \qI$ and $q^i_w = q$, together with canonical transitions $t^1_w, \ldots, t^i_w$; we write $w^j$ for the length-$j$ prefix of $w$, so that $w^j \in \langstate{q^j_w}$.  Note that the canonical run of a prefix is the prefix of the canonical run: $\cpath{w^j}{q^j_w}$ is the length-$j$ prefix of $\cpath{w}{q}$.  The canonical run is a deterministic, purely combinatorial object; the algorithm never computes full canonical runs, but it does test, one layer at a time, whether a given transition is canonical for a given word (Line~\ref{line:canonical-test} of Algorithm~\ref{alg:estsamp}).
}

\subsection{The Algorithm}
\seclabel{algorithm}

$\approxalg$ (Algorithm~\ref{alg:main}) processes the states of $\sautunroll_{n+1}$ in the order $\prec$ and computes, for every state $q$, an estimate $p(q)$ of $\norm{\langstate{q}}^{-1}$ together with $\alpha$ random \emph{sample sets} $S^1(q),\ldots,S^\alpha(q) \subseteq \langstate{q}$, built directly from the sample sets of the predecessors of $q$ by the procedure $\estsamp$ (Algorithm~\ref{alg:estsamp}). The algorithm aims to maintain the (informal) invariant
\begin{equation}
\equlabel{invariant-informal}
\text{``}\Pr[w \in S^r(q)] = p(q)\text{''}
\qquad \text{for every } w \in \langstate{q},
\end{equation}
so that if $p(q)$ is a good estimate of $\norm{\langstate{q}}^{-1}$ then each $S^r(q)$ has constant expected size.  The quoted expression wrongly states that a probability equals a random variable; the rigorous surrogate, $\Ex[\ind{w \in S^r(q)}/p(q)] \approx 1$, is developed in \secref{analysis}.

At each state, we use $\alpha=\beta\gamma$ sample sets, divided into
$\gamma$ blocks of $\beta$ sets each. Each block gives one estimate,
and we take the median of these estimates.
In \secref{analysis}, we show that the relative variance of the estimate
obtained from a block is $O(n/\beta)$. We therefore choose $\beta$
of order $n/\epsl^2$, so that this estimate has relative error
$O(\epsl)$ with probability at least $5/6$. We also show that the joint failure probability of blocks can be controlled
despite their dependence. As a result, the
probability that the median estimate is inaccurate decreases
exponentially in $\gamma$. Taking
$\gamma \in O(\log\norm{\statesunroll})$ then bounds the probability of any
inaccurate state estimate by a small constant, using a union bound over
all states.

The parameters $\thetas$ and $\thetar$ limit the number of letters sampled
from transition predicates and the number of words stored, respectively.
In \lemref{singleton}, we describe the sampling invariant maintained by
the algorithm without its budget checks. This invariant implies that each
sample set contributes only a constant to the expected storage on
executions with accurate estimates, as shown in \lemref{proba-S}.
With $\alpha$ sets per state, this motivates the storage budget
$\thetar \in O(\alpha\norm{\statesunroll})$.
Each stored word can be extended once along each outgoing transition of
its state, which similarly gives the sampling budget
$\thetas \in O(\alpha\norm{\transunroll})$.

For an even number of values, the median is the lower middle value.
The integer choices of $\gamma$ and $\mu$ below satisfy
$\gamma\geq5\ln(16\norm{\statesunroll})$ and $\mu\geq8\ln(1/\dl)$,
the bounds used in the analysis.

\begin{algorithm}[tbp]
\begin{algorithmic}[1]
\State \textbf{if} $n=0$ \textbf{then return} $1$ if $\qI\in\accept$, and $0$ otherwise
\State \textbf{if} $\trans=\emptyset$ \textbf{then return} $0$
\State construct the end-marked automaton and its trimmed unrolling $\sautunroll_{n+1} = (\statesunroll,\transunroll,\qI,\qF)$, memoizing $\acnt{\predphi}$ for every guard; \textbf{if} $\qF$ is trimmed \textbf{then return} $0$ \label{line:preprocess}
\State $\epsl_c = \epsl/4$;\; $\beta = \big\lceil \tfrac{96 (n+2)}{\epsl_c^2 (1-\epsl_c)} \big\rceil$;\; $\gamma = 5\lceil \log_2 (16 \norm{\statesunroll}) \rceil$;\; $\alpha = \beta\gamma$;\; $\thetas = \lceil 32(1{+}\epsl_c) \alpha \norm{\transunroll} \rceil$;\; $\thetar = \lceil 32(1{+}\epsl_c) \alpha \norm{\statesunroll} \rceil$;\; $\mu = 8\lceil \log_2(1/\dl) \rceil$ \label{line:parameters}
\For{$1 \leq j \leq \mu$}
	\State \tikzmark{start}\textbf{for} $q \in \statesunroll$ and $1 \leq r \leq \alpha$ \textbf{do} $p(q) = 1$ and $S^r(q) = \emptyset$;\quad $\ctrs = 0$ \label{line:init}
	\State \textbf{for} $1 \leq r \leq \alpha$ \textbf{do} $S^r(\qI) = \set{\emptyword}$;\quad $\ctrr = \alpha$ \label{line:initS}
	\For{$1 \leq i \leq n+1$}
		\For{$q \in \states^i$ in $\prec$ order}
			\State $\estsamp(q)$ \label{line:call-estsamp}
			\State \textbf{if} $\ctrs = \thetas$ or $\ctrr = \thetar$ \textbf{then} go to Line~\ref{line:endCore} \label{line:abort}
		\EndFor
	\EndFor
	\State \tikzmark{end}$\mathsf{est}_j = 1/p(\qF)$ \label{line:endCore}
\EndFor
\State \Return $\big\lfloor \median(\mathsf{est}_1,\ldots,\mathsf{est}_\mu) + \tfrac12 \big\rfloor$ \label{line:return}
\end{algorithmic}
\caption{$\approxalg(\saut,n,\epsl,\dl)$}
\label{alg:main}
\begin{tikzpicture}[overlay,remember picture]
\draw[thick, decorate, decoration = {calligraphic brace, mirror, amplitude=5pt}] ([xshift=26.5em,yshift=-0.4em]pic cs:end) to node[midway,right] {$\ \ \core$} ([yshift=0.9em,xshift=26.5em]pic cs:start);
\end{tikzpicture}
\end{algorithm}

\begin{algorithm}[tbp]
\begin{algorithmic}[1]
\State $\rho(q) = \minsf\big(p(q_l)/\acnt{\predphi_l} \;\big|\; 1 \leq l \leq k\big)$ \label{line:rho}
\For{$1 \leq r \leq \alpha$ \textbf{and} $1 \leq l \leq k$} \Comment{\footnotesize extend and re-sample\normalsize}
	\State $\Sbar^r(t_l) = \emptyset$
	\For{$v \in S^r(q_l)$}
		\State draw $a = \osample{\predphi_l}$;\; $\ctrs \mathrel{{+}{=}} 1$;\; \textbf{if} $\ctrs = \thetas$ \textbf{then abort} \label{line:letter}
		\State add $v \concat a$ to $\Sbar^r(t_l)$ with probability $\dfrac{\rho(q)\,\acnt{\predphi_l}}{p(q_l)}$ \label{line:extend}
	\EndFor
\EndFor
\For{$1 \leq r \leq \alpha$} \Comment{\footnotesize canonical union\normalsize}
	\State $\Shat^r(q) = \emptyset$
	\For{$1 \leq l \leq k$ \textbf{and} $u = v \concat a \in \Sbar^r(t_l)$}
		\State \textbf{if} for every $l' < l$: $v \notin \langstate{q_{l'}}$ or $\neg\,\omember(a,\predphi_{l'})$ \textbf{then} add $u$ to $\Shat^r(q)$ \label{line:canonical-test}
	\EndFor
\EndFor
\For{$1 \leq b \leq \gamma$}
	\State $Y_{q,b} = \dfrac{1}{\beta\,\rho(q)} \displaystyle\sum_{r = (b-1)\beta + 1}^{b\beta} \norm{\Shat^r(q)}$ \Comment{\footnotesize block means\normalsize} \label{line:blockmean}
\EndFor
\State $\hat\rho(q) = \median(Y_{q,1},\ldots,Y_{q,\gamma})^{-1}$ \Comment{\footnotesize median of means; $0^{-1} = \infty$\normalsize} \label{line:median}
\State $p(q) = \minsf(\rho(q), \hat\rho(q))$; \quad $N(q) = 1/p(q)$ \Comment{\footnotesize $\minsf(x,\infty) = x$\normalsize} \label{line:takemin}
\For{$1 \leq r \leq \alpha$} \Comment{\footnotesize re-sample down to probability $p(q)$\normalsize}
	\State $S^r(q) = \emptyset$;\; \textbf{for} $w \in \Shat^r(q)$ \textbf{do} add $w$ to $S^r(q)$ with probability $p(q)/\rho(q)$, and upon each addition $\ctrr \mathrel{{+}{=}} 1$;\; \textbf{if} $\ctrr = \thetar$ \textbf{then abort} \label{line:finalreduce}
\EndFor
\end{algorithmic}
\caption{$\estsamp(q)$ for $q \in \states^i$, with $\intrans{q} = (t_1 \prec \cdots \prec t_k)$ and $t_l = (q_l,\predphi_l,q)$; the test $v \in \langstate{q_{l'}}$ on Line~\ref{line:canonical-test} is read off the cache computed by (\ref{eq:cache-update}); \textbf{abort} transfers control to Line~\ref{line:abort} of Algorithm~\ref{alg:main}}
\label{alg:estsamp}
\end{algorithm}

The portion of Algorithm~\ref{alg:main} between Lines~\ref{line:init} and~\ref{line:endCore} is called the $\core$; it produces one estimate $\mathsf{est}_j$ of $\norm{\langstate{\qF}} = \numwords{\saut}{n}$, and \secref{analysis} shows that each $\mathsf{est}_j$ lands in $(1 \pm \epsl/2)\numwords{\saut}{n}$ with probability at least $3/4$. Line~\ref{line:return} takes the median of $\mu \in O(\log(1/\dl))$ independent runs and rounds it to the nearest integer, breaking ties upwards. The median amplifies the confidence to $1-\dl$, and the factor-two slack in the error bound ensures that rounding preserves the requested relative error $\epsl$, as shown in the proof of \thmref{symbolic-fpras}. $\core$ essentially processes every state $q$ to generate $p(q)$ and the sets of words $S^1(q),\dots,S^\alpha(q)$. This is done in $\estsamp(q)$. To ensure polynomial running time of $\core$, two counters bound its work: $\ctrs$ counts the sampling calls (Line~\ref{line:letter}) against the budget $\thetas$, and $\ctrr$ counts the stored words (Lines~\ref{line:initS} and~\ref{line:finalreduce}) against the budget $\thetar$.  Each counter is checked as it is incremented, so a run makes at most $\thetas$ draws and stores at most $\thetar$ words; when either budget is reached the run aborts, and \secref{analysis} shows this happens with probability at most $1/8$.  We count an aborted run as a possible failure in the analysis.

%Since every stored word other than the $\alpha$ initial copies of $\emptyword$ descends from the draw of Line~\ref{line:letter} that created it, $\ctrr \leq \ctrs + \alpha$ holds throughout a run; the gap in the other direction is large, because the invariant (\ref{eq:invariant-informal}) keeps each sample set at constant expected size, so a run stores about $\alpha\norm{\statesunroll}$ words while extending each stored word once along each outgoing transition of its state, for about $\alpha\norm{\transunroll}$ draws.  This gap is what the cost accounting of \secref{caching} exploits: a draw is charged $O(\dmax)$ membership bits (one canonical-union test), whereas a stored word is charged a full cache row of $O(\norm{\trans})$ bits, and it is the combination $\thetas\dmax + \thetar\norm{\trans}$---rather than $\thetas(\dmax + \norm{\trans})$---that yields the factor $\dmax + \norm{\states}$ in \thmref{symbolic-fpras}, and hence the cubic dependence on $\norm{\states}$ in Remark~\ref{def:dcm-comparison}.  Replacing the two budgets by a single budget $\max(\thetas,\thetar)$ would leave the algorithm correct but degrade the running time by a factor of $\norm{\trans}/\norm{\states}$.

Now we describe $\estsamp(q)$ for $q \in \states^i$ with incoming transitions $t_1 \prec \cdots \prec t_k$, $t_l = (q_l,\predphi_l,q)$.  At this point, for each predecessor $q_l$ we have computed $p(q_l)$ and constructed the sets $S^r(q_l) \subseteq \langstate{q_l}$ aiming at the informal invariant (\ref{eq:invariant-informal}). The procedure has four phases.

\myparagraph{Extend and reduce (Lines~\ref{line:rho}--\ref{line:extend})}{
In the explicit-alphabet FPRAS of~\cite{arenas2019efficient,decolnet2025practical}, every $v \in S^r(q_l)$ is extended by \emph{every} letter of the (constant-size) alphabet; here $\denote{\predphi_l}$ may be large.  Instead, each $v$ is extended by a \emph{single} letter $a$ drawn by the sampling oracle (Line~\ref{line:letter}), and the $\approx 1/\cnt{\predphi_l}$ probability of drawing any particular letter is compensated by the factor $\acnt{\predphi_l}$ in the retention probability of Line~\ref{line:extend}.  Using the informal invariant, the intended cancellation for any fixed $u = v\concat a \in \langstate{t_l}$ is
\[
p(q_l) \cdot \sdist{\predphi_l}(a) \cdot \frac{\rho(q)\,\acnt{\predphi_l}}{p(q_l)}
\;\approx\; p(q_l) \cdot \frac{1}{\cnt{\predphi_l}} \cdot \frac{\rho(q)\cnt{\predphi_l}}{p(q_l)} \;=\; \rho(q),
\]
So the retention probability of a word in a set $\bar S^r(t_l)$ is around $\rho(q)$ regardless of the incoming transitions. The formal argument is made in \secref{analysis} 

%handles the dependence and oracle bias.  Two points matter.  First, the letter of Line~\ref{line:letter} is fresh for every triple $(r,l,v)$: reusing one letter across the words of $S^r(q_l)$ would correlate distinct words of $\langstate{t_l}$. Second, the retention probability of Line~\ref{line:extend} is at most $1$ precisely because $\rho(q) \leq p(q_l)/\acnt{\predphi_l}$ by Line~\ref{line:rho}.
%defining $\rho(q)$ as a minimum over \emph{transitions}, normalized by guard counts, is the symbolic replacement for the minimum over predecessors of~\cite{decolnet2025practical}, and forcing estimates below $\rho(q)$ is sound because $\norm{\langstate{q}} \geq \norm{\langstate{q_l}}\cnt{\predphi_l}$ by (\ref{eq:transition-product}).
}

\myparagraph{Canonical union (Line~\ref{line:canonical-test})}{
Now we merge the sets $\Sbar^r(t_1),\ldots,\Sbar^r(t_k)$ into one set $\hat S^r(q) \subseteq L(q)$ such that, for any fixed $u \in \langstate{q}$,
\[
\text{``}\Pr\big[u \in \hat S^r(q) \big] = \rho(q)\text{''}.
\]
The sets $\Sbar^r(t_1),\ldots,\Sbar^r(t_k)$ cannot simply be unioned: a word $u \in \langstate{q}$ that reaches $q$ through many incoming transitions would appear with higher probability, biasing the estimator toward words with many derivations, i.e., toward counting runs rather than words. Line~\ref{line:canonical-test} therefore keeps $u = v \concat a$ coming from $\Sbar^r(t_l)$ if and only if $t_l = \canon(u,q)$, i.e., iff no $\prec$-smaller incoming transition $t_{l'}$ validates $u$.  Since $u \in \Sbar^r(t_l)$ is only possible for transitions that validate $u$, and the canonical transition is a deterministic function of $u$, each $u \in \langstate{q}$ survives into $\Shat^r(q)$ with probability $\approx \rho(q)$, through its canonical transition and it alone.  The test $a \sat \predphi_{l'}$ is one oracle call; the test $v \in \langstate{q_{l'}}$ is a cache lookup, using (\ref{eq:cache-update}).

To implement the canonical-transition test, we must decide $v \in \langstate{q_{l'}}$ on Line~\ref{line:canonical-test}.  

Deciding it from scratch is a reachability computation costing $O(\norm{\transunroll})$ oracle calls per test; we instead maintain, layer by layer, a Boolean \emph{cache} matrix, following~\cite{decolnet2025practical}.

Throughout, a word is stored in $O(1)$ space as a tuple (pointer to its length-decremented prefix, last letter), so no word is ever copied.  For $0 \leq i \leq n+1$ let $\calS^i = \bigcup_{r \in [\alpha]}\bigcup_{q \in \states^i} S^r(q)$ denote the set of all words retained for layer-$i$ states.  The cache for layer $i$ is the matrix $\cache_i$ with rows indexed by $\calS^i$ and columns by $\states^i$, and it is \emph{correct} when $\cache_i(w,q) = \ind{w \in \langstate{q}}$ for all entries.  For $i = 0$ the cache is the $1 \times 1$ matrix $(1)$, correct since $\langstate{\qI} = \set{\emptyword}$.  After all states of layer $i$ have been processed, the algorithm computes the row of each retained word $u = v \concat a \in \calS^i$ via the update rule
\begin{equation}
	\equlabel{cache-update}
	\cache_i(v \concat a,\, q) \;=\; \bigvee_{(q',\predpsi,q)\, \in\, \intrans{q}} \cache_{i-1}(v,q') \wedge \omember(a,\predpsi),
\end{equation}
which is correct whenever $\cache_{i-1}$ is, by the definition of runs of $\sautunroll_{n+1}$.  Rows are computed only for retained words: the canonical test of Line~\ref{line:canonical-test} for a candidate $u = v\concat a$ targeted at $q$ reads only the entries $\cache_{i-1}(v, q_{l'})$ of the \emph{previous} layer, which exist since $v$ is retained, together with the bits $\omember(a,\predphi_{l'})$ for $l' < l$.  Membership bits are computed once per pair of a candidate letter and a guard and shared between (\ref{eq:cache-update}) and Line~\ref{line:canonical-test}.

}

\myparagraph{Median-of-means estimate (Lines~\ref{line:blockmean}--\ref{line:takemin})}{ Since each word of $\langstate{q}$ lies in $\Shat^r(q)$ with probability $\approx \rho(q)$, each $\norm{\Shat^r(q)}/\rho(q)$ is an (almost) unbiased estimator of $\norm{\langstate{q}}$; Lines~\ref{line:blockmean}--\ref{line:median} combine the $\alpha = \beta\gamma$ estimates by median-of-means. We take the reciprocal of the median of the block means to obtain $\hat\rho(q)$, an estimate of $1/|L(q)|$. We then set $p(q)=\min(\rho(q),\hat\rho(q))$. To understand this choice, recall that $\rho(q)$ estimates $1/\max_l|L(t_l)|$. Since every $L(t_l)$ is contained in $L(q)$, the target $1/|L(q)|$ is at most $1/\max_l|L(t_l)|$. By taking the minimum, we ensure that $p(q)\leq\rho(q)$, so the final thinning probability $p(q)/\rho(q)$ is at most one.
}

\myparagraph{Reduce (Lines~\ref{line:takemin}--\ref{line:finalreduce})}{ We thin $\Shat^r(q)$ by retaining each word independently with probability $p(q)/\rho(q)$, obtaining $S^r(q)$. Informally, this restores the invariant (\ref{eq:invariant-informal}) at $q$ and keeps the sample sets at constant expected size for use by successor states.
}

The algorithm manipulates only the original guards and draws, stores, and tests concrete letters, each represented in one machine word.

%!TEX root=./main.tex

\section{Analysis}
\seclabel{analysis}
 
We want to prove \thmref{symbolic-fpras}, which guarantees that our algorithm
returns a count within relative error $\epsl$ of the true count with
probability at least $1-\dl$, together with bounds on its oracle queries and
running time.

\subsection*{Notations}
Throughout the section, $\epsl_c = \epsl/4 \in (0,1/4]$ denotes the internal
accuracy parameter fixed on Line~\ref{line:parameters} of
Algorithm~\ref{alg:main}, and $\dtol \leq \epsl_c/(100(n+2)^2)$ is parameter of the $\kappa$-quantitave interface. We work over the trimmed
unrolled automaton $\sautunroll_{n+1} = (\statesunroll, \transunroll, \qI,
\qF)$ of \secref{normalform}, with layers $\states^0,\ldots,\states^{n+1}$.
We abbreviate the original input count by $L := \numwords{\saut}{n}$;
the end-marker construction gives $L = \norm{\langstate{\qF}}$.

%\subsection{Weights, Masses, and the Core Lemmas}
%\seclabel{setup}

\myparagraph{Drift weights}{
	In~\cite{decolnet2025practical} where the task is to approximate the word count of an NFA, $p(q)$ is an estimate of $1/|L(q)|$. Here, on symbolic NFA, because we have to sample from the transition predicates and approximate their sizes, $p(q)$ may drift a bit further away from $1/|L(q)|$. We account for this effect by introducing \emph{drift weights}. Recall from \secref{normalform} that every pair $(w,q)$ with $q \in \states^i$ and $w \in \langstate{q}$ has a unique canonical run $\cpath{w}{q} = (q^0_w,\ldots,q^i_w)$ with canonical transitions $t^1_w,\ldots,t^i_w$, and that the canonical run of a prefix is the prefix of the canonical run. For $1 \leq j \leq i$, writing $t^j_w = (q^{j-1}_w, \predphi_j, q^j_w)$ and $a_j$ for the $j$-th letter of $w$, we define
	   \[
        \lambda_j(w,q) \,:=\, \sdist{\predphi_j}(a_j)\cdot\acnt{\predphi_j},
        \qquad
        \wgt_{>j}(w,q) \,:=\, \prod_{l = j+1}^{i} \lambda_l(w,q),
    \]
    and we abbreviate $\wgt(w,q) := \wgt_{>0}(w,q)$. The $\lambda_j(w,q)$ accounts for the approximate counting and sampling from the predicate $\predphi_j$. 
    
    With those weights, we define the mass of $q$, a value $\mass{q}$ close to $|L(q)|$ (Proposition~\ref{prop:numerics}), whose inverse is actually estimated by $p(q)$.
    $$\mass{q} := \sum_{w \in \langstate{q}} \wgt(w,q).$$ 
    Note that $\mass{\qI} = 1$. Since $\sdist{\predphi}(a) \in [\tfrac{1-\dtol}{\cnt{\predphi}},
            \tfrac{1+\dtol}{\cnt{\predphi}}]$ and $\acnt{\predphi} \in
        [(1-\dtol)\cnt{\predphi}, (1+\dtol)\cnt{\predphi}]$, every $\lambda_j(w,q)$
    satisfies $\lambda_j(w,q) \in [(1-\dtol)^2, (1+\dtol)^2]$, whence for $q \in
        \states^i$:
    \begin{align}
        \equlabel{drift-bounds}
        \wgt_{>j}(w,q) \in \big[(1-\dtol)^{2(i-j)}, (1+\dtol)^{2(i-j)}\big],
        \qquad
        \mass{q} \in \big[(1-\dtol)^{2i}, (1+\dtol)^{2i}\big]\cdot\norm{\langstate{q}}.
    \end{align}
    The drift weights are deterministic: they depend only on the input, the memoized counts, and the fixed distributions of the predicate samplers. 
}

\myparagraph{Target intervals}{
    For $q \in \states^i$ we define the \emph{target interval} $I(q)$, the interval in which we want $p(q)$ to lie, taking the drift phenomenon into account. First, define the constants
    \[
        \driftXi := \left(\frac{1+\dtol}{1-\dtol}\right)^{2(n+2)},
        \qquad
        \tolm_i := \left(\frac{1-\dtol}{1+\dtol}\right)^{i(i+1)} \text{ for } 0 \leq i \leq n+1,
        \qquad
        \tolm := \tolm_{n+1}.
    \]
    Note that $1 = \tolm_0 \geq \tolm_1 \geq \cdots \geq \tolm_{n+1} = \tolm$ and
    $\tolm_i = \tolm_{i-1}\big(\tfrac{1-\dtol}{1+\dtol}\big)^{2i}$. Each $\tolm_i$ is a parameter that captures the consequences at layer $i$ of the \emph{almost}-uniform quality of the samplers for the transition predicates. Now,
    \[
        I(q) := \left[\frac{(1-\epsl_c)\,\tolm_i}{\mass{q}},\; \frac{1+\epsl_c}{\mass{q}}\right].
    \]
    The factor $\tolm_i$ in the lower end comes from \propref{graded-cap}. The target interval $I(q)$ is where we want $p(q)$. Recall that $\rho(q)$ is computed as $p(q')/\acnt{\predphi}$ for some predecessor $q'$ of $q$, with $\predphi$ the guard on the transition. If $p(q')$ is a good estimate $1/|L(q')|$, then $\rho(q)$ is an estimate of $1/(|L(q')|\cdot \cnt{\predphi})$ of slightly worst quality since $\acnt{\predphi}$ gives an approximate value of $\cnt{\predphi}$. The lower quality of the estimate is captured by $\tolm_i$; but is not too dramatic since $\tolm_i \geq 1 - \epsl_c/16 $ by \propref{numerics}(b).
}

The following proposition bounds the effect of oracle error on the analysis.
Part~(a) controls the factor $\driftXi$ that enters the concentration bound.
Part~(b) shows that the cumulative adjustment $\tolm$ to the target intervals
remains close to one, while part~(c) shows that the weighted mass $\mass{q}$
remains close to the true count $|L(q)|$. We use these last two bounds at the
end of the proof to convert the interval guarantee for $p(q)$ into a
relative-error guarantee for the count estimate $1/p(q)$.

\begin{proposition}
	\proplabel{numerics}\leanchecked
	Suppose $\epsl_c \in (0,1/4]$ and $\dtol \leq \epsl_c/(100(n+2)^2)$.  Then
	\begin{itemize}
		\item[(a)] $\driftXi \leq 1 + \epsl_c/8 \leq 2$;
		\item[(b)] $\tolm \geq 1 - \epsl_c/16 \geq 1/2$;
		\item[(c)] $(1-\epsl_c/16)\norm{\langstate{q}} \leq \mass{q} \leq (1+\epsl_c/16)\norm{\langstate{q}}$ for every $q \in \statesunroll$.
	\end{itemize}
\end{proposition}
\begin{proof}
	We prove (a) and (b) by taking logarithms and applying the assumed bound
	on $\dtol$. For (c), we apply elementary estimates to the upper and lower
	bounds on the weighted mass.
	
	Our assumptions imply $\dtol \leq 1/3$, so
	\[
	\ln\frac{1+\dtol}{1-\dtol}
	\leq \frac{2\dtol}{1-\dtol}
	\leq 3\dtol.
	\]
	We will also use $e^x \leq 1+2x$ for $0 \leq x \leq 1/2$, and
	$e^{-x} \geq 1-x$ for $x \geq 0$.
	
	\emph{(a)} We take the logarithm of $\driftXi$ and substitute the bound
	on $\dtol$:
	\[
	\ln\driftXi
	= 2(n+2)\ln\frac{1+\dtol}{1-\dtol}
	\leq 6(n+2)\dtol
	\leq \frac{6\epsl_c}{100(n+2)}
	\leq \frac{\epsl_c}{16}.
	\]
	Since $\epsl_c/16 \leq 1/2$, we obtain
	$\driftXi \leq e^{\epsl_c/16} \leq 1+\epsl_c/8 \leq 2$.
	
	\emph{(b)} By the definition of $\tolm = \tolm_{n+1}$, we have
	\[
	\begin{aligned}
		\ln(1/\tolm)
		&= (n+1)(n+2)\ln\frac{1+\dtol}{1-\dtol} \\
		&\leq 3(n+2)^2\dtol
		\leq \frac{3\epsl_c}{100}
		\leq \frac{\epsl_c}{16}.
	\end{aligned}
	\]
	Applying $e^{-x} \geq 1-x$, we conclude that
	$\tolm \geq e^{-\epsl_c/16} \geq 1-\epsl_c/16 \geq 1/2$.
	
	\emph{(c)} Fix a state $q$ in layer $i \leq n+1$. We first bound the
	accumulated error along a word reaching $q$:
	\[
	2i\dtol
	\leq \frac{2(n+1)\epsl_c}{100(n+2)^2}
	\leq \frac{\epsl_c}{50}
	\leq \frac{\epsl_c}{32}.
	\]
	For the upper bound, we use (\ref{eq:drift-bounds}) and
	$1+\dtol \leq e^{\dtol}$:
	\[
	\frac{\mass{q}}{\norm{\langstate{q}}}
	\leq (1+\dtol)^{2i}
	\leq e^{2i\dtol}
	\leq e^{\epsl_c/32}
	\leq 1+\frac{\epsl_c}{16}.
	\]
	For the lower bound, we use Bernoulli's inequality:
	\[
	\frac{\mass{q}}{\norm{\langstate{q}}}
	\geq (1-\dtol)^{2i}
	\geq 1-2i\dtol
	\geq 1-\frac{\epsl_c}{32}
	\geq 1-\frac{\epsl_c}{16}.
	\]
	Multiplying both bounds by $\norm{\langstate{q}}$ proves (c).
\end{proof}

\subsection*{Proof Outline}

 To establish accuracy, we must
therefore address two possible sources of failure: \textbf{(a) inaccurate
	estimates and (b) early termination} when one of these budgets is exhausted.

We remove the budget checks to obtain $\corestar$. This is the same algorithm as $\core$ but which does not stop too many samples are drawn, so $\core$ and $\corestar$
behave identically until a budget is reached. We conduct the analysis of the estimates accuracy in $\corestar$(\lemref{proba-p}), and use this to show that the probability of exceeding the budgets is small
(\lemref{proba-S}). Thus, analyzing $\corestar$, we obtain a guarantee for one run of the actual algorithm $\core$ (\lemref{core-accuracy}). Finally, we obtain the required confidence by taking the median of independent runs and rounding it to the nearest integer.

To address (a), we must control how errors propagate when each state uses
estimates and samples produced at earlier states. We introduce an auxiliary
algorithm $\corespirit$ in which an benign spirit corrects estimates whenever they leave their prescribed accuracy ranges, which turns out to be very helpful in the proof. By
showing that spirited algorithm $\corespirit$ agrees with $\corestar$ until the first spirit intervention, we reduce the task of bounding errors in $\corestar$ to bounding
the probability that the spirit intervenes in $\corespirit$ (\lemref{to-spirit}). We can then
analyze a single state in $\corespirit$ assuming that the preceding estimates satisfy their
bounds (\lemref{rho-hat}).

To establish accuracy at that state, we must show that the median of the block
estimates is unlikely to be inaccurate. Since the blocks share earlier random
choices, we cannot treat their errors as independent. We instead bound the
probability that several blocks are inaccurate simultaneously
(\lemref{joint-looseness}). We obtain this bound by tracing the relevant joint
moments backwards through the layers using conditional-expectation identities
(\lemref{induction-descent}).
%, and reducing the remaining calculation to a
%combinatorial bound on pairs of words (\lemref{divergence-mass}). We account
%for collisions explicitly: when two words require different letters from the
%same draw, they cannot both be sampled.

We must also account for the bias introduced by approximate guard operations.
We do this by assigning each word a drift weight describing its accumulated bias and
analyzing the resulting weighted count $\mass{q}$. We have already shown that this weighted count
remains close to the true count (\propref{numerics}).

\subsection{The Uninterrupted Algorithm}

We write $\corestar$ for one run of $\core$ except that the two counters (hence
the abort) are removed. $\core$ and $\corestar$ behave the same until the number of letter draws or the number of stored words grow too big (beyond their respective budget). In $\corestar$, define
\[
    D := \sum_{q \in \statesunroll}\sum_{r \in [\alpha]} \norm{S^r(q)}\cdot\norm{\outtrans{q}},
    \qquad
    R := \sum_{q \in \statesunroll}\sum_{r \in [\alpha]} \norm{S^r(q)}.
\]
$D$ is the total number of draws in a run of $\corestar$ and $R$ the
total number of stored words; in $\core$, the counters $\ctrs$ and $\ctrr$
trace exactly these quantities, and both are non-decreasing over a run. The two
core lemmas of the analysis are the following.

\begin{lemma}
    \lemlabel{proba-p}\leanchecked
    In $\corestar$, $\Pr\left[\,\exists q \in \statesunroll:\ p(q) \notin I(q)\,\right] \leq \dfrac{1}{16}$.
\end{lemma}

\begin{lemma}
    \lemlabel{proba-S}\leanchecked
    In $\corestar$, $\Pr\left[\,D \geq \thetas \ \text{or}\ R \geq \thetar\,\right] \leq \dfrac{1}{8}$.
\end{lemma}

The first says that, in $\corestar$, each estimate lays in its target interval with good probability. The second says that $\corestar$ is unlikely to go over-budget, and thus is likely to behave like $\core$. Therefore the first lemma also applies to $\core$ with a modified probability. Thus granting these lemmas, the correctness of one run of $\core$ follows.

\begin{lemma}
	\lemlabel{core-accuracy}\leanchecked
	In $\core$, $\Pr\left[\mathsf{est}_j \notin (1 \pm \frac{\epsl}{2})\,L\right] \leq \dfrac{1}{4}$, where $\epsl = 4\epsl_c$.
\end{lemma}

We are now ready to prove \thmref{symbolic-fpras}.
\begin{proof}[Proof of \thmref{symbolic-fpras}]
	The cases $n=0$ and $\trans=\emptyset$ are handled exactly in preprocessing.
	If $\qF$ is trimmed, the output $0$ is also exact. Otherwise
%	, we first
%	account for the time limit on generating retention decisions. The proof
%	of \lemref{core-accuracy} bounds the probability of leaving the target
%	interval by $1/8+1/16=3/16$ and gives relative error at most $\epsl/2$
%	whenever $p(\qF)\in I(\qF)$.
%	We couple each implemented run to the ideal run using the same random choices. They
%	agree until the time limit is reached, which occurs with probability at
%	most $1/16$ by the proof of \lemref{core-cost}. Hence each implemented run fails with
%	probability at most $3/16+1/16=1/4$.
	the different runs to get $\mathsf{est}_1,\ldots,\mathsf{est}_\mu$ use independent randomness, and a standard amplification argument is used to argue that their median has
	relative error at most $\epsl/2$ with probability at least $1-\dl$.
	Write $M=\median(\mathsf{est}_1,\ldots,\mathsf{est}_\mu)$ for this median,
	let $X_j$ indicate that $\mathsf{est}_j\notin(1\pm\epsl/2)L$, and set
	$\bar X=X_1+\cdots+X_\mu$. \lemref{core-accuracy} gives
	$\Ex[\bar X]\leq\mu/4$. If $M$ is outside $(1\pm\epsl/2)L$, at least
	$\mu/2$ of the estimates are outside, so Hoeffding's inequality
	(\propref{concentration-tools}(b)) gives
	\[
	\Pr\left[M \notin \left(1\pm\frac{\epsl}{2}\right)L\right]
	\leq \Pr\left[\bar X - \Ex[\bar X] \geq \tfrac{\mu}{4}\right]
	\leq e^{-\mu/8} \leq \dl,
	\]
	using $\mu \geq 8\ln(1/\dl)$.
	The algorithm returns $\widetilde N = \lfloor M+\tfrac12\rfloor$.
	Since $\widetilde N$ is a nearest integer to $M$ and $L$ is an integer,
	$|\widetilde N-M| \leq |L-M|$. Consequently, whenever
	$|M-L| \leq \epsl L/2$,
	\[
	|\widetilde N-L| \leq |\widetilde N-M|+|M-L|
	\leq 2|M-L| \leq \epsl L.
	\]
	Thus the nonnegative integer output satisfies the required approximation
	guarantee with probability at least $1-\dl$.
	
	For the cost, \lemref{core-cost} bounds
	one implemented run of $\core$ by $\thetas$ sampling calls and $O(\thetas\dmax +
	\thetar\norm{\trans})$ membership calls and word operations, on top of
	the $O(\norm{\trans} + \norm{\states})$ memoized counting calls of
	preprocessing. In the trimmed unrolling, $\norm{\statesunroll} \leq
	(n+2)(\norm{\states}+1)$, and $\norm{\transunroll} \leq (n+1)\norm{\trans} +
	\norm{\accept}$, where the second addend counts the end-marker transitions;
	every accepting state that survives trimming is reachable, hence has an
	incoming transition or equals $\qI$, so $\norm{\accept} \leq \norm{\trans}+1$
	and $\norm{\transunroll} \in O(n\norm{\trans})$. With $\beta \in O(n\,\epsl^{-2})$
	and $\gamma \in O(\log(n\norm{\states}))$,
	\[
	\thetas \in O(\alpha\norm{\transunroll}) \subseteq O\!\left(n^2\,\norm{\trans}\,\log(n\norm{\states})\,\epsl^{-2}\right),
	\qquad
	\thetar \in O(\alpha\norm{\statesunroll}) \subseteq O\!\left(n^2\,\norm{\states}\,\log(n\norm{\states})\,\epsl^{-2}\right);
	\]
	multiplying by the $\mu \in O(\log(1/\dl))$ runs and by the per-call costs
	$\Tsamp$ and $\Tmem$ gives the running time of \thmref{symbolic-fpras}, and
	setting $\Tcnt, \Tsamp, \Tmem \in O(1)$ gives \corref{unit-cost}.
	The estimates have the common denominator $\beta^{n+1}$, including the
	value $1$ returned on a numerical timeout. Selecting the final median
	therefore costs $O(\mu n)$ word operations on their integer numerators.
	Computing the common denominator and rounding the median by integer
	division cost $O(n^2)$ word operations. These costs fit within the bound
	above, since a nonempty unrolling has at least $n+1$ transitions and
	$\alpha=\Omega(n)$. Neither step makes oracle queries.
\end{proof}

The rest of the section is organized as follows. \secref{spirit} introduces the
spirited algorithm and reduces \lemref{proba-p} to a concentration bound for
the median-of-means statistic. \secref{environment} sets up normalized word contributions
and their conditional-expectation identities. \secref{concentration} proves the concentration
bound and \secref{sample-size} proves \lemref{proba-S}.

\subsection{The Spirited Algorithm}
\seclabel{spirit}

We introduce the algorithm $\corespirit$, which follows $\corestar$ but is
helped by an omniscient benign spirit. In $\corespirit$, the value computed on
Line~\ref{line:takemin} of $\estsamp(q)$ is renamed $\bar p(q) :=
    \minsf(\rho(q), \hat\rho(q))$; if $\bar p(q) \in I(q)$ the spirit stays quiet
and $p(q) = \bar p(q)$, and otherwise the spirit intervenes and sets
\[
    p(q) := \frac{(1-\epsl_c)\,\tolm_i}{\mass{q}} \qquad (q \in \states^i),
\]
i.e., the lower end of $I(q)$. Thus in $\corespirit$ the guarantee $p(q) \in
    I(q)$ holds with probability $1$ for every processed state. We first verify
that the spirit's replacement value keeps all retention and thinning
probabilities in $[0,1]$.

\begin{proposition}
    \proplabel{graded-cap}\leanchecked
    In $\corespirit$, for every $q \in \states^i$ with $i \geq 1$,
    $\rho(q) \geq \dfrac{(1-\epsl_c)\tolm_i}{\mass{q}}$.
\end{proposition}
\begin{proof}
    Let $t = (q',\predphi,q) \in \intrans{q}$ attain the minimum on Line~\ref{line:rho}, so $\rho(q) = p(q')/\acnt{\predphi}$ with $q' \in \states^{i-1}$.  The spirit guarantees $p(q') \geq (1-\epsl_c)\tolm_{i-1}/\mass{q'}$, hence
    $\rho(q) \geq \tfrac{(1-\epsl_c)\tolm_{i-1}}{\mass{q'}\acnt{\predphi}}$.
    By (\ref{eq:drift-bounds}), (\ref{eq:transition-product}), and $\langstate{t} \subseteq \langstate{q}$,
    \begin{align*}
        \mass{q'}\cdot\acnt{\predphi}
         & \;\leq\; (1+\dtol)^{2(i-1)}\norm{\langstate{q'}}\cdot(1+\dtol)\cnt{\predphi}
        \;=\; (1+\dtol)^{2i-1}\norm{\langstate{t}}                                      \\
         & \;\leq\; (1+\dtol)^{2i-1}\norm{\langstate{q}}
        \;\leq\; \left(\frac{1+\dtol}{1-\dtol}\right)^{2i}\mass{q},
    \end{align*}
    and therefore $\rho(q) \geq (1-\epsl_c)\tolm_{i-1}\big(\tfrac{1-\dtol}{1+\dtol}\big)^{2i}/\mass{q} = (1-\epsl_c)\tolm_i/\mass{q}$.
\end{proof}

In particular, the spirit's replacement value is at most $\rho(q)$, so the
thinning probability $p(q)/\rho(q)$ of Line~\ref{line:finalreduce} remains in
$[0,1]$, as do the retention probabilities of Line~\ref{line:extend} by the
definition of $\rho$. The next proposition shows that, in $\corespirit$, the
only way $\bar p(q)$ can leave $I(q)$ is through the median-of-means statistic.

\begin{proposition}
    \proplabel{min-preserves}\leanchecked
    In $\corespirit$, if $\hat\rho(q) \in \frac{1 \pm \epsl_c}{\mass{q}}$ then $\bar p(q) \in I(q)$.
\end{proposition}
\begin{proof}
    If $\bar p(q) = \hat\rho(q)$ the claim is immediate since $\tfrac{1-\epsl_c}{\mass{q}} \geq \tfrac{(1-\epsl_c)\tolm_i}{\mass{q}}$.  Otherwise $\bar p(q) = \rho(q) < \hat\rho(q)$; then $\bar p(q) \leq \hat\rho(q) \leq \tfrac{1+\epsl_c}{\mass{q}}$, and $\bar p(q) = \rho(q) \geq \tfrac{(1-\epsl_c)\tolm_i}{\mass{q}}$ by \propref{graded-cap}.
\end{proof}

We now transfer the failure event from $\corestar$ to $\corespirit$ by a
coupling on execution prefixes, exactly as in~\cite{decolnet2025practical}. For
a fixed input, an \emph{execution} of $\corestar$ (resp.\ $\corespirit$) is an
assignment of outcomes to all its random choices (letter draws, retention
coins, thinning coins), and a \emph{$q$-prefix execution} assigns exactly the
choices made for states $\preceq q$. While the spirit has not intervened,
$\corestar$ and $\corespirit$ make the same choices with the same
probabilities, so a $q$-prefix execution in which all states $q' \prec q$ have
$p(q') \in I(q')$ has the same probability under both algorithms.

\begin{lemma}
    \lemlabel{to-spirit}\leanchecked
    $\displaystyle
        \Pr_{\corestar}\left[\exists q:\ p(q) \notin I(q)\right]
        \;\leq\; \sum_{q \in \statesunroll \setminus \set{\qI}} \Pr_{\corespirit}\left[\hat\rho(q) \notin \tfrac{1\pm\epsl_c}{\mass{q}}\right].
    $
\end{lemma}
\begin{proof}
    If the event on the left occurs, there is a $\prec$-least state $q$ with $p(q) \notin I(q)$; this state is not $\qI$ since $p(\qI) = 1 = \mass{\qI}^{-1} \in I(\qI)$.  Hence
    \[
        \Pr_{\corestar}\left[\exists q:\, p(q) \notin I(q)\right]
        = \sum_{q \succ \qI} \Pr_{\corestar}\left[p(q) \notin I(q) \text{ and } \forall q' \prec q:\, p(q') \in I(q')\right].
    \]
    Fix $q$. The event in the summand is a union of $q$-prefix executions of
    $\corestar$ in which no state before $q$ is bad; each such prefix execution is
    also a $q$-prefix execution of $\corespirit$ in which the spirit intervenes for
    the first time at $q$ (i.e., $\bar p(q) \notin I(q)$ and $\bar p(q') = p(q')$
    for all $q' \prec q$), and it carries the same probability under both
    algorithms, since the two algorithms use the same coins while the spirit is
    quiet. Therefore
    \begin{align*}
         & \Pr_{\corestar}\left[p(q) \notin I(q),\ \forall q' {\prec} q: p(q') \in I(q')\right]                    \\
         & \qquad = \Pr_{\corespirit}\left[\bar p(q) \notin I(q),\ \forall q' {\prec} q: \bar p(q') = p(q')\right]
        \;\leq\; \Pr_{\corespirit}\left[\bar p(q) \notin I(q)\right],
    \end{align*}
    and $\Pr_{\corespirit}[\bar p(q) \notin I(q)] \leq \Pr_{\corespirit}[\hat\rho(q) \notin \tfrac{1\pm\epsl_c}{\mass{q}}]$ by \propref{min-preserves}.
\end{proof}

In view of \lemref{to-spirit}, \lemref{proba-p} reduces to the following
per-state concentration bound, whose proof occupies \secref{environment} and
\secref{concentration}.

\begin{lemma}
    \lemlabel{rho-hat}\leanchecked
    In $\corespirit$, for every $q \in \statesunroll \setminus \set{\qI}$,
    $\Pr\left[\hat\rho(q) \notin \tfrac{1\pm\epsl_c}{\mass{q}}\right] \leq \dfrac{1}{16\norm{\statesunroll}}$.
\end{lemma}

\subsection{Conditional Expectations of Sample Contributions}
\seclabel{environment}

We analyze the samples by conditioning on the history before each random step.
We can then treat the sampling probabilities for that step as fixed. We
normalize the word contributions so that their conditional expectations equal
the contributions from the preceding step. We then show that, by repeatedly
applying the tower rule, we can reduce expectations at a later layer to
quantities at layer $0$. We use two histories per layer: one before the sampled
extensions and one before the final thinning.

All statements in this subsection hold in both $\corespirit$ and $\corestar$.
The proofs use the sampling and thinning rules, without requiring the bounds
on $p(q)$ supplied by the spirit. For uniform notation, set $\bar p(q)=p(q)$
in $\corestar$. We use the tower rule, the fixed-variable rule, and
conditional independence from \secref{prelim}.

\myparagraph{Sampling histories}{
    For $0 < i \leq n+1$, let $\calF_i$ contain all random variables associated
    with states in layers $< i$. Let $\hat{\calF}_i$ additionally contain the
    letter draws and retention coins of
    Lines~\ref{line:letter}--\ref{line:extend} for all layer-$i$ states.
    Thus $\hat{\calF}_i$ fixes the sets $\Sbar^r(t)$ and $\Shat^r(q)$ at
    layer $i$, before final thinning. We set $\calF_0 = \emptyset$ and use
    the conventions $\Shat^r(\qI) := \set{\emptyword}$ and $\rho(\qI) := 1$.

    For $q \in \states^i$, the value $\rho(q)$ is fixed by $\calF_i$, while
    $\hat\rho(q)$, $\bar p(q)$, and $p(q)$ are fixed by $\hat{\calF}_i$.
    The thinning coins of Line~\ref{line:finalreduce} are then included in
    $\calF_{i+1}$. These histories satisfy
    $\calF_i \subset \hat{\calF}_i \subset \calF_{i+1}$.
}

\myparagraph{Normalized word contributions}{
    We track each word's contribution to a sample set along its canonical run.
    Dividing its membership indicator by the current sampling parameter
    cancels the retention or thinning factor when we take conditional
    expectations. We also multiply by the weight $\wgt_{>j}(w,q)$ to account for the
    oracle factors on the remaining transitions.
    Fix $q \in \states^i$ and $w \in \langstate{q}$. For $r \in [\alpha]$
    and $0 \leq j \leq i$, define
    \[
        A_r(w,q,j) := \wgt_{>j}(w,q)\,\frac{\ind{w^j \in S^r(q^j_w)}}{p(q^j_w)},
        \qquad
        \hat A_r(w,q,j) := \wgt_{>j}(w,q)\,\frac{\ind{w^j \in \Shat^r(q^j_w)}}{\rho(q^j_w)},
    \]
    and their centered versions $B_r(w,q,j) := A_r(w,q,j) - \wgt(w,q)$ and $\hat
        B_r(w,q,j) := \hat A_r(w,q,j) - \wgt(w,q)$. 
    Since $p(q)$ and $\rho(q)$ depend on earlier random choices, the informal
    sampling invariant from \secref{approx} must be stated in terms of these
    normalized contributions. We will show in \corref{atom-expectation} that
    their expectation is $\wgt(w,q)$ at every layer. With exact oracles this
    weight is $1$; with approximate oracles it records the accumulated bias.

    At the top layer the product
    $\wgt_{>i}(w,q)$ is empty: $\hat A_r(w,q,i) = \ind{w \in \Shat^r(q)}/\rho(q)$,
    and since $\Shat^r(q) \subseteq \langstate{q}$ (immediate induction: sample
    sets only ever contain words with a run to their state),
    \begin{align}
        \equlabel{Y-decomposition}
        Y_{q,b} \;=\; \frac{1}{\beta}\sum_{r \in R_b}\frac{\norm{\Shat^r(q)}}{\rho(q)}
        \;=\; \frac{1}{\beta}\sum_{r \in R_b}\sum_{w \in \langstate{q}} \hat A_r(w,q,i),
    \end{align}
    where $R_b = \set{(b-1)\beta+1, \ldots, b\beta}$ is the $b$-th block.
    At layer $0$, we have
    \[
        A_r(w,q,0) = \wgt(w,q)\ind{\emptyword \in S^r(\qI)}/p(\qI)
        = \wgt(w,q)
    \]
    deterministically, so $B_r(w,q,0) = 0$.
}

\begin{lemma}
    \lemlabel{singleton}\leanchecked
    Let $q \in \states^i$, $w \in \langstate{q}$, $r \in [\alpha]$, and $1 \leq j \leq i$.  Then
    \[
        \Ex[A_r(w,q,j) \mid \hat{\calF}_j] = \hat A_r(w,q,j)
        \qquad\text{and}\qquad
        \Ex[\hat A_r(w,q,j) \mid \calF_j] = A_r(w,q,j-1).
    \]
\end{lemma}
\begin{proof}
    Write $t^j_w = (c, \predphi, q^j_w)$ with $c = q^{j-1}_w$, let $a$ be the $j$-th letter of $w$, and abbreviate $v = w^{j-1}$, so $w^j = v \concat a$.

    For the first identity: $p(q^j_w)$ and $\rho(q^j_w)$ are fixed by
    $\hat{\calF}_j$, and $w^j \in S^r(q^j_w)$ iff $w^j$ survives the thinning of
    Line~\ref{line:finalreduce} applied to $\Shat^r(q^j_w)$ with probability
    $p(q^j_w)/\rho(q^j_w)$. Since $\ind{w^j \in \Shat^r(q^j_w)}$ is fixed by
    $\hat{\calF}_j$ and the coin is independent of $\hat{\calF}_j$,
    \[
        \Ex[A_r(w,q,j) \mid \hat{\calF}_j]
        = \frac{\wgt_{>j}(w,q)}{p(q^j_w)}\cdot\frac{p(q^j_w)}{\rho(q^j_w)}\,\ind{w^j \in \Shat^r(q^j_w)}
        = \hat A_r(w,q,j).
    \]
    For the second identity: by the canonical-union test of
    Line~\ref{line:canonical-test}, $w^j$ can enter $\Shat^r(q^j_w)$ only through
    its canonical transition $t^j_w$, and $w^j \in \Sbar^r(t^j_w)$ iff (1) $v \in
        S^r(c)$, (2) the fresh letter drawn for the triple $(r, t^j_w, v)$ on
    Line~\ref{line:letter} equals $a$, and (3) the retention coin of
    Line~\ref{line:extend} succeeds. Given $\calF_j$, event (1) is fixed, event (2)
    has probability $\sdist{\predphi}(a)$, and event (3) has probability
    $\rho(q^j_w)\acnt{\predphi}/p(c)$; events (2) and (3) are independent given
    $\calF_j$. Hence
    \[
        \Ex[\hat A_r(w,q,j) \mid \calF_j]
        = \frac{\wgt_{>j}(w,q)}{\rho(q^j_w)}\cdot \sdist{\predphi}(a)\cdot\frac{\rho(q^j_w)\acnt{\predphi}}{p(c)}\,\ind{v \in S^r(c)}
        = \wgt_{>j}(w,q)\,\lambda_j(w,q)\,\frac{\ind{v \in S^r(c)}}{p(c)},
    \]
    which is $A_r(w,q,j-1)$ since $\wgt_{>j}(w,q)\lambda_j(w,q) =
        \wgt_{>j-1}(w,q)$.
\end{proof}

\begin{corollary}
    \corlabel{atom-expectation}\leanchecked
    For all $q \in \states^i$, $w \in \langstate{q}$, $r \in [\alpha]$, and $0 \leq j \leq i$: $\Ex[A_r(w,q,j)] = \Ex[\hat A_r(w,q,j)] = \wgt(w,q)$.  In particular $\Ex[Y_{q,b}] = \mass{q}$.
\end{corollary}
\begin{proof}
    Taking expectations in the two identities of \lemref{singleton} and
    applying the tower rule gives
    \[
        \Ex[A_r(w,q,j)] = \Ex[\hat A_r(w,q,j)] = \Ex[A_r(w,q,j-1)].
    \]
    Repeating this equality down to layer $0$, where both contributions
    equal $\wgt(w,q)$ deterministically, proves the first claim.
    The claim on $Y_{q,b}$ follows by linearity from (\ref{eq:Y-decomposition}).
\end{proof}

\myparagraph{Divergence and collisions}{
Fix $q \in \states^i$ and $w_1, w_2 \in \langstate{q}$.  Since the canonical run of a prefix is the prefix of the canonical run, the set of indices $j$ with $(w_1^j, q^j_{w_1}) = (w_2^j, q^j_{w_2})$ is downward closed; we define the \emph{divergence level} $\Div(w_1,w_2)$ as its maximum, and the \emph{divergence state} $\divv{w_1}{w_2} := q^{k}_{w_1}$ where $k = \Div(w_1,w_2)$.  We have $\Div(w_1,w_2) = i$ iff $w_1 = w_2$ (in which case $\divv{w_1}{w_2} = q$), and $\Div(w_1,w_2) \geq 0$ always, since layer $0$ is shared by all runs.

We now describe the collision phenomenon. Say $w_1 \neq w_2$ with $k =
    \Div(w_1,w_2)$. We call the pair \emph{colliding} when $t^{k+1}_{w_1} =
    t^{k+1}_{w_2}$, i.e., when the two canonical runs leave their last common point
through the \emph{same} transition; in that case the $(k{+}1)$-th letters of
$w_1, w_2$ necessarily differ. For a colliding pair, the words $w_1^{k+1}$ and
$w_2^{k+1}$ both descend from the single letter drawn for the triple $(r,
    t^{k+1}_{w_1}, w_1^k)$, so they can never enter $\Sbar^r(t^{k+1}_{w_1})$
together. }

For example, suppose that the only transition into $q_2$ is $(q_1,\predphi,q_2)$, where $\denote{\predphi}=\{0,1\}$.
Fix a repetition $r$ in which $00\in S^r(q_1)$.
To extend this prefix along the transition, we draw a single letter from $\predphi$: drawing $0$ gives $000$, while drawing $1$ gives $001$.
Either word may subsequently be retained, but they cannot both enter $S^r(q_2)$, since this would require the same draw to return two different letters.
Thus $000$ and $001$ form a colliding pair, with divergence state $q_1$.

\begin{lemma}
    \lemlabel{pair}\leanchecked
    Let $q \in \states^i$, $w_1, w_2 \in \langstate{q}$ with $(w_1^j, q^j_{w_1}) \neq (w_2^j, q^j_{w_2})$ (i.e., $j > k := \Div(w_1,w_2)$), and $r \in [\alpha]$.  Then
    \[
        \Ex[A_r(w_1,q,j)A_r(w_2,q,j) \mid \hat{\calF}_j] = \hat A_r(w_1,q,j)\hat A_r(w_2,q,j),
    \]
    and
    \[
        \begin{aligned}
            \Ex[\hat A_r(w_1,q,j)\hat A_r(w_2,q,j) \mid \calF_j] =
            \begin{cases}
                0 \hfill \text{if } j - 1 = k,\ (w_1,w_2) \text{ colliding}, \\[2pt]
                A_r(w_1,q,j-1)\,A_r(w_2,q,j-1) \qquad\qquad\qquad\hfill \text{otherwise.}
            \end{cases}
        \end{aligned}
    \]
    Moreover, for $r_1 \neq r_2$ and any $w_1, w_2$, the two products descend with
    no collision case:
    \begin{align*}
        \Ex[\hat A_{r_1}(w_1,q,j)\hat A_{r_2}(w_2,q,j) \mid \calF_j]
         & = A_{r_1}(w_1,q,j-1)A_{r_2}(w_2,q,j-1),       \\
        \Ex[A_{r_1}(w_1,q,j)A_{r_2}(w_2,q,j) \mid \hat{\calF}_j]
         & = \hat A_{r_1}(w_1,q,j)\hat A_{r_2}(w_2,q,j).
    \end{align*}
\end{lemma}
\begin{proof}
    For the first identity: the pairs $(w_1^j, q^j_{w_1}) \neq (w_2^j, q^j_{w_2})$ are distinct, so the two events $w_l^j \in S^r(q^j_{w_l})$ are governed by distinct thinning coins of Line~\ref{line:finalreduce}, which are mutually independent given $\hat{\calF}_j$; the probabilities $p(\cdot)/\rho(\cdot)$ of these coins are fixed by $\hat{\calF}_j$.  The computation of \lemref{singleton} then applies to each contribution separately.

    For the second identity, condition on $\calF_j$ and write $t_l := t^j_{w_l}$,
    $v_l := w_l^{j-1}$ and $a_l$ for the $j$-th letter of $w_l$ ($l = 1,2$). As in
    \lemref{singleton}, $w_l^j \in \Shat^r(q^j_{w_l})$ iff $v_l \in
        S^r(q^{j-1}_{w_l})$, the letter drawn for $(r, t_l, v_l)$ equals $a_l$, and the
    retention coin succeeds. If $(t_1, v_1) \neq (t_2, v_2)$, the letter draws and
    coins involved are distinct, hence mutually independent given $\calF_j$, and
    the product factors into the two singleton computations, yielding
    $A_r(w_1,q,j-1)A_r(w_2,q,j-1)$. If $(t_1,v_1) = (t_2,v_2)$, then the two events
    require the \emph{same} letter draw to equal both $a_1$ and $a_2$; the two
    pairs $(w_1^j, q^j_{w_1}) \neq (w_2^j, q^j_{w_2})$ share their state (the
    target of $t_1$) and their prefix $v_1 = v_2$, so $a_1 \neq a_2$, the joint
    event is impossible, and the conditional expectation is $0$. It remains to
    observe that $(t_1,v_1) = (t_2,v_2)$ means precisely that the canonical runs
    agree at level $j-1$ and leave through the same transition, i.e., $j - 1 = k$
    and the pair is colliding.

    For $r_1 \neq r_2$, the two contributions use disjoint randomness given $\calF_j$
    (resp.\ $\hat{\calF}_j$), and conditional independence together with
    \lemref{singleton} gives the claim.
\end{proof}

\myparagraph{Correction terms for pairs of words}{
    To bound the squared error of a block, we must control products of two
    word contributions. By \lemref{pair}, their conditional expectation
    changes when they share a membership indicator or require different
    letters from the same draw. We introduce correction terms that account for these
    cases. They will give an upper bound on the squared error whose
    expectation we can evaluate one layer at a time.

    Fix $q \in \states^i$ and a pair $(w_1,w_2) \in \langstate{q}^2$ with $k = \Div(w_1,w_2)$, and set
    \[
        \Phi(w_1,w_2) := \frac{\mass{\divv{w_1}{w_2}}}{(1-\epsl_c)\,\tolm}.
    \]
    In $\corespirit$, $\Phi(w_1,w_2)$ bounds the inverse sampling parameter
    at the divergence state. We use this bound in \secref{concentration};
    the identities below hold in both algorithms.
    For $r \in [\alpha]$ and $0 \leq j \leq i$, define
    \[
        \begin{aligned}
            \hat C_r(w_1,w_2,q,j) :=
            \begin{cases}
                \Phi(w_1,w_2)\, p(\divv{w_1}{w_2})\, \hat A_r(w_1,q,j) - \wgt(w_1,q) & \text{if } j > k,                                      \\[3pt]
                \Phi(w_1,w_2)\, \wgt_{>k}(w_1,q) - \wgt(w_1,q)                       & \text{if } j \leq k,\ (w_1,w_2) \text{ not colliding}, \\[3pt]
                -\,\wgt(w_1,q)                                                       & \text{if } j \leq k,\ (w_1,w_2) \text{ colliding},
            \end{cases}
        \end{aligned}
    \]
    and $C_r(w_1,w_2,q,j)$ identically with $\hat A_r$ replaced by $A_r$.
    The first case applies after divergence, when the contributions use
    distinct membership indicators and $p(\divv{w_1}{w_2})$ is fixed by
    $\calF_j$. For a non-colliding pair, the two indicators coincide at
    layer $k$; the factor $p(\divv{w_1}{w_2})$ then cancels one inverse
    sampling factor in their product, giving the second case. For a colliding
    pair, the joint extension event at layer $k+1$ is impossible, leaving
    only the negative term in the third case. The next lemma verifies these
    identities.

    Diagonal pairs $w_1=w_2$ have $k=i$, $\wgt_{>i}(w_1,q)=1$, and are never
    colliding. Their correction term is therefore the constant
    $\Phi(w_1,w_1)-\wgt(w_1,q)$ at every layer.
}

\begin{lemma}
    \lemlabel{C-descent}\leanchecked
    For $q \in \states^i$, $(w_1,w_2) \in \langstate{q}^2$, $r \in [\alpha]$, and $1 \leq j \leq i$:
    \[
        \Ex[C_r(w_1,w_2,q,j)\,A_r(w_2,q,j) \mid \hat{\calF}_j] = \hat C_r(w_1,w_2,q,j)\,\hat A_r(w_2,q,j),
    \]
    \[
        \Ex[\hat C_r(w_1,w_2,q,j)\,\hat A_r(w_2,q,j) \mid \calF_j] = C_r(w_1,w_2,q,j-1)\,A_r(w_2,q,j-1).
    \]
\end{lemma}
\begin{proof}
    Let $k = \Div(w_1,w_2)$.  We consider three regimes.

    \emph{Case $j - 1 \geq k+1$, or $w_1 = w_2$.}  If $w_1 = w_2$ the $C$-term is a constant and both identities are instances of \lemref{singleton}.  Otherwise $j > j-1 > k$: the $C$-terms at levels $j$ and $j-1$ are both in the first case, and $p(\divv{w_1}{w_2})$ is fixed by $\calF_j$ (hence by $\hat{\calF}_j$).  Expanding,
    \[
        \hat C_r(w_1,w_2,q,j)\hat A_r(w_2,q,j) = \Phi\,p(\divv{w_1}{w_2})\,\hat A_r(w_1,q,j)\hat A_r(w_2,q,j) - \wgt(w_1,q)\,\hat A_r(w_2,q,j),
    \]
    and since $j - 1 > k$ the pair is not at its merge level, so \lemref{pair} (no
    collision) and \lemref{singleton} descend the two summands exactly, giving
    $C_r(w_1,w_2,q,j-1)A_r(w_2,q,j-1)$. The $\hat{\calF}_j$-direction is identical
    using the first parts of \lemref{pair} and \lemref{singleton}.

    \emph{Case $j - 1 = k$ (merge level), $w_1 \neq w_2$.}  The $\hat{\calF}_j$-direction is as before.  For the $\calF_j$-direction, suppose first the pair is \emph{not} colliding.  By \lemref{pair},
    \[
        \Ex[\hat A_r(w_1,q,j)\hat A_r(w_2,q,j) \mid \calF_j] = A_r(w_1,q,k)\,A_r(w_2,q,k).
    \]
    At level $k$ the two canonical prefixes coincide: $w_1^k = w_2^k$ and
    $q^k_{w_1} = q^k_{w_2} = \divv{w_1}{w_2}$, so
    \[
        A_r(w_1,q,k)A_r(w_2,q,k)
        = \wgt_{>k}(w_1,q)\wgt_{>k}(w_2,q)\frac{\ind{w_1^k \in S^r(\divv{w_1}{w_2})}}{p(\divv{w_1}{w_2})^2}
        = \frac{\wgt_{>k}(w_1,q)}{p(\divv{w_1}{w_2})}\,A_r(w_2,q,k).
    \]
    Hence
    \begin{align*}
         & \Ex[\hat C_r(w_1,w_2,q,j)\hat A_r(w_2,q,j) \mid \calF_j]                     \\
         & \qquad = \big(\Phi(w_1,w_2)\,\wgt_{>k}(w_1,q) - \wgt(w_1,q)\big)A_r(w_2,q,k)
        = C_r(w_1,w_2,q,k)\,A_r(w_2,q,k),
    \end{align*}
    as required, the $p(\divv{w_1}{w_2})$ having cancelled.  If instead the pair \emph{is} colliding, then \lemref{pair} gives $\Ex[\hat A_r(w_1,q,j)\hat A_r(w_2,q,j) \mid \calF_j] = 0$, so only the second summand survives:
    \[
        \Ex[\hat C_r(w_1,w_2,q,j)\hat A_r(w_2,q,j) \mid \calF_j]
        = -\wgt(w_1,q)\,A_r(w_2,q,k)
        = C_r(w_1,w_2,q,k)\,A_r(w_2,q,k).
    \]

    \emph{Case $j - 1 < k$, $w_1 \neq w_2$.}  Then both the $C$-terms for levels $j$ and $j-1$ are in the constant regime ($j \leq k$), the constant is the same at both levels, and both identities reduce to instances of \lemref{singleton}.
\end{proof}

\subsection{Proof of \lemref{rho-hat} and \lemref{proba-p}} \seclabel{concentration}

We work in $\corespirit$ and fix $q \in \states^i \setminus \set{\qI}$. Call
block $b \in [\gamma]$ \emph{failing} when $Y_{q,b} \notin
    (1\pm\epsl_c)^{-1}\mass{q} := \big[\tfrac{\mass{q}}{1+\epsl_c},
        \tfrac{\mass{q}}{1-\epsl_c}\big]$ and \emph{loose} when $\norm{Y_{q,b} -
        \mass{q}} \geq \tfrac{\epsl_c}{2}\mass{q}$. Since $0 < \epsl_c \leq 1$ we have
$(1 \pm \tfrac{\epsl_c}{2})\mass{q} \subseteq (1\pm\epsl_c)^{-1}\mass{q}$, so
every failing block is loose. If $\hat\rho(q) \notin
    \tfrac{1\pm\epsl_c}{\mass{q}}$ then the median of $Y_{q,1},\ldots,Y_{q,\gamma}$
is outside $(1\pm\epsl_c)^{-1}\mass{q}$, which requires at least $\gamma/2$
failing, hence loose, blocks:
\begin{align}
    \equlabel{rho-to-blocks}
    \Pr\left[\hat\rho(q) \notin \tfrac{1\pm\epsl_c}{\mass{q}}\right]
    \leq \Pr\left[\sum_{b \in [\gamma]} \ind{\norm{Y_{q,b} - \mass{q}} \geq \frac{\epsl_c}{2}\mass{q}} \geq \frac{\gamma}{2}\right].
\end{align}
The blocks are not independent, so we use the following elementary surrogate for the Chernoff bound.

\begin{proposition}
    \proplabel{chernoff-intersections}\leanchecked
    Let $E_1,\ldots,E_\gamma$ be events such that $\Pr[\bigcap_{b \in F} E_b] \leq x^{\norm{F}}$ for every $F \subseteq [\gamma]$, where $x \in [0,1]$.  Then $\Pr\left[\sum_b \ind{E_b} \geq \gamma/2\right] \leq 2^\gamma x^{\gamma/2}$.
\end{proposition}
\begin{proof}
    If at least $\gamma/2$ of the events occur, then $\bigcap_{b \in F} E_b$ occurs for some $F$ of size $\lceil\gamma/2\rceil$.  A union bound over the at most $2^\gamma$ such sets, each of probability at most $x^{\lceil \gamma/2\rceil} \leq x^{\gamma/2}$, completes the proof.
\end{proof}

It therefore remains to bound the probability that several blocks are loose
\emph{simultaneously}.

\begin{lemma}
    \lemlabel{joint-looseness}\leanchecked
    For every $F \subseteq [\gamma]$,
    \[
        \Pr_{\corespirit}\left[\bigcap_{b \in F} \norm{Y_{q,b} - \mass{q}} \geq \tfrac{\epsl_c}{2}\mass{q}\right]
        \;\leq\;
        \left(\frac{4\,(n+2)\,\driftXi}{\epsl_c^2 (1-\epsl_c)\,\tolm\,\beta}\right)^{\norm{F}}
        \;\leq\; \frac{1}{6^{\norm{F}}}.
    \]
\end{lemma}

Before proving \lemref{joint-looseness}, we first derive \lemref{rho-hat} and
\lemref{proba-p} from it. By (\ref{eq:rho-to-blocks}) and
\propref{chernoff-intersections} with $x = 1/6$,
\[
    \Pr\left[\hat\rho(q) \notin \tfrac{1\pm\epsl_c}{\mass{q}}\right]
    \leq 2^\gamma\, 6^{-\gamma/2} = \left(\tfrac{2}{3}\right)^{\gamma/2}
    \leq e^{-\gamma/5}
    \leq \frac{1}{16\norm{\statesunroll}}
\]
using $\gamma \geq 5\ln(16\norm{\statesunroll})$, which proves
\lemref{rho-hat}. Summing over the at most $\norm{\statesunroll}$ states in
\lemref{to-spirit} proves \lemref{proba-p}. \hfill$\square$

\subsubsection{Proof of \lemref{joint-looseness}}

So far, we have shown that each block estimate has mean $\mass{q}$. We now
need to bound the probability that several blocks are inaccurate
simultaneously. We first use Markov's inequality to bound this probability in
terms of the joint moment
\[
    \Ex\!\left[\prod_{b \in F}(Y_{q,b} - \mass{q})^2\right].
\]
We expand the squared errors and use the correction terms to obtain an upper
bound. At each step, we condition on the shared
history and use the independence of fresh random choices across blocks to
factor the conditional expectation of the product. The preceding lemmas
express each factor in terms of the previous layer. Repeating this calculation
down to layer $0$ leaves a deterministic quantity, which we bound
combinatorially.

\emph{Step 1: Markov.}  Since the events intersect, by \propref{concentration-tools}(a),
\begin{align}
    \Pr\left[\bigcap_{b \in F} \norm{Y_{q,b} - \mass{q}} \geq \tfrac{\epsl_c}{2}\mass{q}\right] & \leq \Pr\left[\prod_{b \in F} (Y_{q,b} - \mass{q})^2 \geq \left(\tfrac{\epsl_c \mass{q}}{2}\right)^{2\norm{F}}\right] \notag \\
                                                                                                & \quad \leq \left(\tfrac{2}{\epsl_c \mass{q}}\right)^{2\norm{F}} \Ex\left[\prod_{b \in F} (Y_{q,b} - \mass{q})^2\right].
    \equlabel{markov-step}
\end{align}

\emph{Step 2: decomposition and domination.}  Fix a block $b$.  By (\ref{eq:Y-decomposition}) and $\mass{q} = \sum_w \wgt(w,q)$, we have $\beta^2(Y_{q,b} - \mass{q})^2 = (\sum_{r \in R_b}\sum_{w} \hat B_r(w,q,i))^2$, which can be written
\begin{align}
    \beta^2 (Y_{q,b} - \mass{q})^2
     & = \sum_{\substack{r \in R_b         \\ w_1,w_2}} \Big(\hat A_r(w_1,q,i)\hat A_r(w_2,q,i) - \wgt(w_1,q)\hat A_r(w_2,q,i)\Big) \notag \\
     & \quad -\; \sum_{\substack{r \in R_b \\ w_1,w_2}} \wgt(w_2,q)\hat B_r(w_1,q,i) + \sum_{\substack{r_1 \neq r_2 \in R_b\\ w_1,w_2}} \hat B_{r_1}(w_1,q,i)\hat B_{r_2}(w_2,q,i),
    \equlabel{square-decomposition}
\end{align}
where $w_1, w_2$ range over $\langstate{q}$.  We now dominate the first sum.  For $w_1 \neq w_2$ let $k = \Div(w_1,w_2)$, the spirit guarantees $p(\divv{w_1}{w_2}) \geq (1-\epsl_c)\tolm_k/\mass{\divv{w_1}{w_2}} \geq (1-\epsl_c)\tolm/\mass{\divv{w_1}{w_2}}$, i.e., $\Phi(w_1,w_2)\,p(\divv{w_1}{w_2}) \geq 1$; since the product of contributions is non-negative,
\begin{align*}
    \hat A_r(w_1,q,i)\hat A_r(w_2,q,i) - \wgt(w_1,q)\hat A_r(w_2,q,i)
     & \;\leq\; \hat C_r(w_1,w_2,q,i)\,\hat A_r(w_2,q,i).
\end{align*}
For $w_1 = w_2 =: w$, using $\ind{}^2 = \ind{}$ and \propref{graded-cap} (which holds deterministically in $\corespirit$),
\[
    \hat A_r(w,q,i)^2 = \frac{\hat A_r(w,q,i)}{\rho(q)} \leq \frac{\mass{q}}{(1-\epsl_c)\tolm}\,\hat A_r(w,q,i) = \Phi(w,w)\,\hat A_r(w,q,i),
\]
so the same domination holds for the diagonal pairs $(w,w)$. Defining, for $0
    \leq j \leq i$,
\[
    \hat N_{b,j} := \sum_{\substack{r \in R_b\\ w_1,w_2}} \hat C_r(w_1,w_2,q,j)\,\hat A_r(w_2,q,j),
\]
\[
    \hat V_{b,j} := \sum_{\substack{r_1 \neq r_2 \in R_b\\ w_1,w_2}} \hat B_{r_1}(w_1,q,j)\hat B_{r_2}(w_2,q,j) - \sum_{\substack{r \in R_b\\ w_1,w_2}} \wgt(w_2,q)\hat B_r(w_1,q,j),
\]
and $N_{b,j}, V_{b,j}$ identically with unhatted contributions and terms, we have shown
\begin{align}
    \equlabel{dominated}
    0 \leq \beta^2 (Y_{q,b} - \mass{q})^2 \leq \hat N_{b,i} + \hat V_{b,i}.
\end{align}

\emph{Step 3: conditioning through the layers.}

\begin{lemma}
    \lemlabel{induction-descent}\leanchecked
    For every $1 \leq j \leq i$,
    \[
        \begin{aligned}
            \Ex\left[\prod_{b \in F} (N_{b,j} + V_{b,j})\right] = \Ex\left[\prod_{b \in F} (\hat N_{b,j} + \hat V_{b,j})\right] = \Ex\left[\prod_{b \in F} (N_{b,j-1} + V_{b,j-1})\right].
        \end{aligned}
    \]
\end{lemma}
\begin{proof}
    For the second identity, condition on $\calF_j$. This fixes the weights,
    the layer-$j$ values $\rho(\cdot)$, and the values
    $p(\divv{w_1}{w_2})$ for pairs with $\Div(w_1,w_2)<j$.
    Thus, only the membership indicators in $\hat N_{b,j}+\hat V_{b,j}$
    remain random. These depend on fresh letter draws and retention coins.
    Distinct blocks use disjoint repetitions, so their fresh choices are
    independent given $\calF_j$. Consequently,
    \[
        \Ex\left[\prod_{b \in F} (\hat N_{b,j} + \hat V_{b,j}) \,\Big|\, \calF_j\right] = \prod_{b \in F} \Ex\left[\hat N_{b,j} + \hat V_{b,j} \mid \calF_j\right].
    \]
    By \lemref{C-descent}, $\Ex[\hat N_{b,j}\mid\calF_j]=N_{b,j-1}$.
    Similarly, \lemref{singleton} and the cross-repetition identities of
    \lemref{pair} give $\Ex[\hat V_{b,j}\mid\calF_j]=V_{b,j-1}$:
    subtracting the constant $\wgt(w,q)$ preserves the conditional-expectation
    identities for $B$ and $\hat B$. Substituting these equalities and taking
    expectations proves the second identity by the tower rule.

    For the first identity, condition instead on $\hat{\calF}_j$. This also
    fixes the layer-$j$ values $p(\cdot)$, leaving only the final thinning
    coins, which are independent across repetitions. Factor over blocks as
    above and apply the $\hat{\calF}_j$ identities of
    Lemmas~\ref{lem:C-descent}, \ref{lem:singleton}, and~\ref{lem:pair}.
    Taking expectations proves the first identity.
\end{proof}

Alternating the two identities of \lemref{induction-descent} from layer $i$
down to layer $0$ yields
\begin{align}
    \equlabel{down-to-zero}
    \Ex\left[\prod_{b \in F} (\hat N_{b,i} + \hat V_{b,i})\right] = \prod_{b \in F} (N_{b,0} + V_{b,0}),
\end{align}
the right-hand side being deterministic.  Since $B_r(w,q,0) = 0$, we get $V_{b,0} = 0$: the cross-repetition and linear sums vanish, as promised.

\emph{Step 4: evaluation at layer $0$.}  At layer $0$ every $C$-term is in its constant regime ($j = 0 \leq \Div(w_1,w_2)$ always), and $A_r(w_2,q,0) = \wgt(w_2,q)$.  Bounding the colliding constants by $0$ and the non-colliding constants by $\Phi(w_1,w_2)\wgt_{>\Div(w_1,w_2)}(w_1,q)$,
\begin{align}
    N_{b,0}
     & \;=\; \sum_{\substack{r \in R_b                                                                                                             \\ w_1,w_2}} C_r(w_1,w_2,q,0)\,\wgt(w_2,q) \notag \\
     & \;\leq\; \frac{\beta}{(1-\epsl_c)\tolm} \sum_{w_1,w_2 \in \langstate{q}} \mass{\divv{w_1}{w_2}}\,\wgt_{>\Div(w_1,w_2)}(w_1,q)\,\wgt(w_2,q).
    \equlabel{N-zero}
\end{align}
The double sum is controlled by the following weighted divergence bound, the symbolic counterpart of the derivation-path argument of~\cite{decolnet2025practical}.

\begin{lemma}
    \lemlabel{divergence-mass}\leanchecked
    For every $q \in \states^i$,
    \[
        \sum_{w_1,w_2 \in \langstate{q}} \mass{\divv{w_1}{w_2}}\,\wgt_{>\Div(w_1,w_2)}(w_1,q)\,\wgt(w_2,q) \;\leq\; (n+2)\,\driftXi\,\mass{q}^2.
    \]
\end{lemma}

Let us first consider exact guard oracles (in which case all weights are one).
For a fixed $w_1$ and $j$, we want to bound $|\langstate{q_j}|$ times the number of words that diverge from $w_1$ at level $j$.

Consider a deterministic layered automaton in which every run to $q$ at layer $4$ passes through $r$ at layer $2$.
Suppose $\langstate{r}=\{00,01,10\}$, and the possible continuations from $r$ to $q$ are $00,10,11$.
Thus $\langstate{q}=\{0000,0010,0011,0100,0110,0111,1000,1010,1011\}$ contains 9 words.

For $w_1=0000$ and $j=2$, the divergence state is $r$, and the words under consideration are $0010$ and $0011$.
Both begin with $00$.
We may replace this common prefix by any of the three words in $\langstate{r}$: after reading the replacement prefix, we are again at $r$, from where the original suffix still leads to $q$.
This gives three choices of prefix and two choices of suffix.
Since every prefix has length two, different choices give different words, and hence
\[
    |\langstate{r}|\cdot|\{0010,0011\}|=3\cdot2\leq |\langstate{q}|=9.
\]
In the proof below, we make the same argument for each fixed $w_1$ and $j$.
With approximate oracles, we must additionally compare the weights before and after splicing.
In a nondeterministic automaton, the concatenated run may differ from the canonical run of the resulting word; the ratio of their weights is bounded by $\driftXi$.

\begin{proof}
    Fix $w_1 \in \langstate{q}$ and $0 \leq j \leq i$, and let $\Dset{w_1}{j} := \setpred{w_2 \in \langstate{q}}{\Div(w_1,w_2) = j}$; write $q_j := q^j_{w_1}$.  Every $w_2 \in \Dset{w_1}{j}$ satisfies $w_2^j = w_1^j$ and $q^j_{w_2} = q_j$, so by prefix-consistency of canonical runs, $\wgt(w_2,q) = \wgt(w_1^j, q_j)\cdot\wgt_{>j}(w_2,q)$, and we may write $w_2 = w_1^j \concat s(w_2)$ with $s(w_2)$ the length-$(i-j)$ suffix of $w_2$.  Hence
    \begin{align*}
         & \sum_{w_2 \in \Dset{w_1}{j}} \mass{q_j}\,\wgt_{>j}(w_1,q)\,\wgt(w_2,q) = \wgt_{>j}(w_1,q)\,\wgt(w_1^j,q_j) \sum_{x \in \langstate{q_j}} \sum_{w_2 \in \Dset{w_1}{j}} \wgt(x,q_j)\,\wgt_{>j}(w_2,q),
    \end{align*}
    and $\wgt_{>j}(w_1,q)\,\wgt(w_1^j,q_j) = \wgt(w_1,q)$.  Consider now the map $(x, w_2) \mapsto y := x \concat s(w_2)$ on $\langstate{q_j} \times \Dset{w_1}{j}$.  The letters of $s(w_2)$ satisfy the guards of the canonical transitions $t^{j+1}_{w_2},\ldots,t^{i}_{w_2}$, which form a path of $\sautunroll_{n+1}$ from $q_j$ to $q$; since $x \in \langstate{q_j}$, we get $y \in \langstate{q}$.  The map is injective, because $x = y^j$ and $w_2 = w_1^j \concat s(w_2)$ are recovered from $y$.  Finally, both $\wgt(x,q_j)\wgt_{>j}(w_2,q)$ and $\wgt(y,q)$ are products of exactly $i$ values, each lying in $[(1-\dtol)^2,(1+\dtol)^2]$, so their ratio is at most $\big(\tfrac{1+\dtol}{1-\dtol}\big)^{2i} \leq \driftXi$.  Therefore
    \[
        \sum_{x,\,w_2} \wgt(x,q_j)\,\wgt_{>j}(w_2,q)
        \;\leq\; \driftXi \sum_{x,\,w_2} \wgt\big(x \concat s(w_2),\, q\big)
        \;\leq\; \driftXi\,\mass{q}
    \]
    by injectivity. Summing over the $i+1 \leq n+2$ values of $j$ and then over
    $w_1$,
    \begin{align*}
         & \sum_{w_1}\sum_{j = 0}^{i}\; \sum_{w_2 \in \Dset{w_1}{j}} \mass{q_j}\,\wgt_{>j}(w_1,q)\,\wgt(w_2,q) \\
         & \qquad \;\leq\; \sum_{w_1} \wgt(w_1,q)\,(n+2)\,\driftXi\,\mass{q}
        \;=\; (n+2)\,\driftXi\,\mass{q}^2. \qedhere
    \end{align*}
\end{proof}

We now return to the proof of \lemref{joint-looseness} and combine the bounds
established above.

\emph{Step 5: collecting.}  By (\ref{eq:dominated}), each factor of $\prod_{b \in F} \beta^2(Y_{q,b} - \mass{q})^2$ is non-negative and pointwise dominated by $\hat N_{b,i} + \hat V_{b,i}$, so the product is pointwise dominated as well.  Combining this with (\ref{eq:down-to-zero}), $V_{b,0} = 0$, (\ref{eq:N-zero}), and \lemref{divergence-mass},
\begin{align*}
    \Ex\left[\prod_{b \in F} (Y_{q,b} - \mass{q})^2\right]
     & \leq \frac{1}{\beta^{2\norm{F}}}\,\Ex\left[\prod_{b \in F}(\hat N_{b,i} + \hat V_{b,i})\right]
    = \frac{1}{\beta^{2\norm{F}}}\prod_{b \in F} N_{b,0}                                              \\
     & \leq \left(\frac{(n+2)\,\driftXi\,\mass{q}^2}{(1-\epsl_c)\,\tolm\,\beta}\right)^{\norm{F}}.
\end{align*}
Substituting into (\ref{eq:markov-step}),
\[
    \Pr\left[\bigcap_{b \in F} \norm{Y_{q,b} - \mass{q}} \geq \tfrac{\epsl_c}{2}\mass{q}\right]
    \leq \left(\frac{4\,(n+2)\,\driftXi}{\epsl_c^2(1-\epsl_c)\,\tolm\,\beta}\right)^{\norm{F}}
    \leq \left(\frac{16\,(n+2)}{\epsl_c^2(1-\epsl_c)\,\beta}\right)^{\norm{F}}
    \leq \frac{1}{6^{\norm{F}}},
\]
using $\driftXi \leq 2$ and $\tolm \geq 1/2$ (\propref{numerics}) and then
$\beta = \lceil 96(n+2)/(\epsl_c^2(1-\epsl_c))\rceil$. This proves
\lemref{joint-looseness}. \hfill$\square$

\subsection{Proof of \lemref{proba-S}}
\seclabel{sample-size}

We work in $\corestar$. By \lemref{proba-p} and a union bound,
\begin{align*}
    \Pr_{\corestar}\left[D \geq \thetas \text{ or } R \geq \thetar\right]
     & \leq \frac{1}{16}
    + \Pr_{\corestar}\left[D\prod_{q'}\ind{p(q') \in I(q')} \geq \thetas\right] + \Pr_{\corestar}\left[R\prod_{q'}\ind{p(q') \in I(q')} \geq \thetar\right],
\end{align*}
since on the complement of the bad event of \lemref{proba-p} all the indicators equal $1$.  We bound the expectation of one summand of $R$ under the indicator at its own state: on the event $p(q) \in I(q)$ we have $1 \leq \tfrac{1+\epsl_c}{\mass{q}\,p(q)}$, and $\norm{S^r(q)}/p(q) = \sum_{w \in \langstate{q}} A_r(w,q,i)$, so by \corref{atom-expectation} (which holds in $\corestar$),
\[
    \Ex\left[\norm{S^r(q)}\,\ind{p(q) \in I(q)}\right]
    \leq \frac{1+\epsl_c}{\mass{q}}\,\Ex\left[\frac{\norm{S^r(q)}}{p(q)}\right]
    = \frac{1+\epsl_c}{\mass{q}}\sum_{w \in \langstate{q}} \wgt(w,q)
    = 1 + \epsl_c.
\]
Dropping all indicators except the one at $q$ and applying Markov bound to each
of the two terms,
\begin{align*}
    \Pr_{\corestar} & \left[D\prod_{q'}\ind{p(q') \in I(q')} \geq \thetas\right]
    \leq \frac{1}{\thetas} \Ex\left[D\prod_{q'}\ind{p(q') \in I(q')}\right]
    \\
                    & \leq \frac{1}{\thetas}\sum_{q}\norm{\outtrans{q}}\sum_{r} \Ex\left[\norm{S^r(q)}\,\ind{p(q) \in I(q)}\right]
    \leq \frac{(1+\epsl_c)\,\alpha\,\norm{\transunroll}}{\thetas}
    \leq \frac{1}{32},
\end{align*}
and likewise $\Pr_{\corestar}[R\prod_{q'}\ind{p(q') \in I(q')} \geq \thetar] \leq (1+\epsl_c)\alpha\norm{\statesunroll}/\thetar \leq 1/32$, using the budgets $\thetas = \lceil 32(1+\epsl_c)\alpha\norm{\transunroll}\rceil$ and $\thetar = \lceil 32(1+\epsl_c)\alpha\norm{\statesunroll}\rceil$ of Line~\ref{line:parameters}.  Summing, $\frac{1}{16} + \frac{1}{32} + \frac{1}{32} = \frac{1}{8}$. \hfill$\square$

\subsection{The Cost of $\core$}
\seclabel{caching}

The following lemma bounds the cost of one execution of $\core$. 
\begin{lemma}
	\lemlabel{core-cost}
	The word-RAM implementation of one run of $\core$ makes at most $\thetas$ calls to $\osample{\cdot}$ and $O(\thetas\dmax + \thetar\norm{\trans})$ calls to $\omember$, and performs $O(\thetas\dmax + \thetar\norm{\trans})$ additional word operations.
\end{lemma}
\begin{proof}
	The counter $\ctrs$ caps the number of draws at $\thetas$, and each draw creates at most one candidate; the counter $\ctrr$ caps the number of stored words at $\thetar$.  Each candidate $u = v\concat a$ targeted at $q$ is represented by a pointer and a letter, using $O(1)$ word operations, and processed by Line~\ref{line:canonical-test} using at most $\norm{\intrans{q}}$ membership bits and cache lookups. We account for the random retention decisions below.  For $q$ in layers $1$ through $n$, $\norm{\intrans{q}} \leq \dmax$; the candidates targeted at $\qF$ number at most $\thetar$ (one per accepting state and word stored at layer $n$) and incur $\norm{\intrans{\qF}} = \norm{\accept} \leq \norm{\trans}+1$ bits each, a total of $O(\thetar\norm{\trans})$.  Each retained word receives a cache row via (\ref{eq:cache-update}) at a cost of at most $\norm{\trans}$ membership bits and $O(\norm{\trans})$ Boolean operations.  This gives $\thetas$ sampling calls and $O(\thetas\dmax + \thetar\norm{\trans})$ membership calls and Boolean operations.  Computing the block totals and selecting their median takes $O(\alpha)$ word operations per state, hence $O(\thetar)$ in total.
	
	To account for larger numbers, represent each estimate $N(q)=1/p(q)$ at
	layer $i$ with the common denominator $\beta^i$. To compute $1/\rho(q)$,
	multiply each predecessor's numerator by the corresponding guard count and
	take the largest product. The numerator of $N(q)$ is this product multiplied
	by the larger of $\beta$ and the median block total. Each block total is at
	most $\thetas$, so every multiplying factor occupies a constant number of
	machine words. Each layer therefore adds only a constant number of words
	to the numerator, which occupies $O(i)$ words. We compute each transition
	product once and reuse it, including in the retention probabilities. These
	comparisons and multiplications cost
	$O(n(\norm{\transunroll}+\norm{\statesunroll}))$ per run, hence $O(\thetas)$,
	since $\alpha=\Omega(n)$ and
	$\norm{\statesunroll}\leq\norm{\transunroll}+1$.
	
	We implement the random retention decisions exactly using cached integer
	numerators and denominators. We store each denominator without leading
	zero words and record its length. For each decision, generate a candidate
	with the same length: choose its leading word uniformly from zero through
	the denominator's leading word, and its remaining words uniformly.
	Retain below the numerator, discard at or above the numerator but below
	the denominator, and retry otherwise. Generate candidate words from most to least
	significant, stopping as soon as the comparisons determine the outcome;
	shorter numerators are read with leading zeros, without copying them.
	At most two candidate prefixes leave the outcome undecided, so the
	probability of reading further words decreases geometrically. Each attempt
	therefore takes constant expected time, and at least half of the candidate
	range lies below the denominator. Thus each decision takes constant
	expected time. For the leading word, use a full random word when all word
	values are allowed; otherwise, rejection using integer division samples
	the required range in constant expected time.
	
	A run makes at most $2\thetas$ such decisions: one after each draw and at
	most one when thinning each surviving candidate. We cap their total work,
	including all rejection attempts, at a sufficiently large constant multiple
	of $\thetas$. If this limit is reached, the implementation stops the run
	and returns $1$. Markov's inequality bounds this probability by $1/16$.
	The implementation agrees with the displayed algorithm until this limit
	is reached; the proof of \thmref{symbolic-fpras} accounts for the additional failure probability.
	The ideal $\core$ analyzed in \secref{analysis} retains only the two budgets
	shown in the pseudocode.
\end{proof}

Substituting $\thetas \in O(\alpha\norm{\transunroll}) \subseteq O\!\big(n^2\allowbreak \norm{\trans}\allowbreak\log(n \norm{\states})\,\allowbreak\epsl^{-2}\big)$ and $\thetar \in O(\alpha\norm{\statesunroll}) \subseteq O\!\big(n^2\allowbreak \norm{\states}\allowbreak\log(n \norm{\states})\,\allowbreak\epsl^{-2}\big)$, and multiplying by the $\mu \in O(\log(1/\dl))$ runs of $\core$, yields the cost claimed in \thmref{symbolic-fpras} and \corref{unit-cost}.

\section{Proof of \lemref{core-accuracy}}

\begin{proof}	
		$\core$ and $\corestar$ follow identical distributions until $\ctrs$ reaches $\thetas$ or $\ctrr$ reaches $\thetar$.  If neither happens, the runs coincide; since $D$ and $R$ are the terminal values of these counters in $\corestar$,
	\[
	\Pr_{\core}\left[p(\qF) \notin I(\qF)\right]
	\leq \Pr_{\corestar}\left[D \geq \thetas \text{ or } R \geq \thetar\right]
	+ \Pr_{\corestar}\left[p(\qF) \notin I(\qF)\right]
	\leq \frac{1}{8} + \frac{1}{16} = \frac{3}{16}.
	\]
	When $p(\qF) \in I(\qF)$, the estimate $\mathsf{est}_j = 1/p(\qF)$ satisfies
	\[
	\frac{\mass{\qF}}{1+\epsl_c}
	\;\leq\; \mathsf{est}_j
	\;\leq\; \frac{\mass{\qF}}{(1-\epsl_c)\,\tolm}.
	\]
	
	By \propref{numerics}(b,c), $\tolm \geq 1 - \epsl_c/16$ and $\mass{\qF} \in (1
	\pm \epsl_c/16)L$. For the upper bound, using $\epsl_c \leq 1/4$,
	\[
	\mathsf{est}_j \leq \frac{(1+\epsl_c/16)\,L}{(1-\epsl_c)(1-\epsl_c/16)}
	\leq \frac{1+\epsl_c/16}{1 - \tfrac{17}{16}\epsl_c}\,L
	\leq (1+2\epsl_c)\,L,
	\]
	where the last step follows from
	\[
	(1+2\epsl_c)(1-\tfrac{17}{16}\epsl_c)
	-(1+\tfrac{\epsl_c}{16})
	= \frac{\epsl_c(7-17\epsl_c)}{8} \geq 0,
	\]
	since $\epsl_c \leq 1/4$. For the lower bound,
	\[
	\mathsf{est}_j \geq \frac{(1-\epsl_c/16)L}{1+\epsl_c} \geq (1-\epsl_c/16)(1-\epsl_c)\,L \geq \left(1 - \tfrac{17}{16}\epsl_c\right) L \geq (1 - 2\epsl_c)\,L.
	\]
	Thus, since $2\epsl_c = \epsl/2$, the proof gives the stronger bound
	\begin{equation}
		\equlabel{core-half-accuracy}
		\Pr_{\core}\left[\mathsf{est}_j \notin (1 \pm \epsl/2)\,L\right]
		\leq \frac14,
	\end{equation}
	which in particular implies the stated bound.
\end{proof}

%!TEX root=./main.tex

\section{Related Work}
\seclabel{related}

\myparagraph{Symbolic automata}
A rich literature develops the theory, algorithms, and applications of
symbolic automata~\cite{dantoni2017power,veanes2021automata}. It includes
language operations and equivalence checking~\cite{veanes2010rex,veanes2012symbolic},
minimization~\cite{dantoni2014minimization},
bisimulation~\cite{dantoni2017bisimulations}, and
active learning~\cite{drews2017learning}. Applications range from
reasoning about string sanitizers~\cite{HooimeijerLMSV11} to
regular-expression matching~\cite{saarikivi2019symbolic,moseley2023derivative}.
We build on the separation between finite-state structure and reasoning
about transition predicates. Our counting algorithm uses membership
tests, approximate counts, and almost-uniform samples for the input
guards, without requiring Boolean operations on predicates.

\myparagraph{Approximate counting for formal languages}
Early work gave quasi-polynomial algorithms for counting and sampling
in regular~\cite{kannan1995counting} and
context-free~\cite{gore1997sampling} languages. Arenas et
al.~\cite{arenas2019efficient,arenas2021nfa} established the first
FPRAS for \cntNFA{}, followed by improved bounds~\cite{meel2024faster,decolnet2025practical}
and the \textsc{FastNFA} implementation~\cite{heissl2026fastnfa}.
FPRAS results also cover nondeterministic tree
automata~\cite{arenas2021tree}, context-free
grammars~\cite{meel2026cfg}, nondeterministic ordered binary decision
diagrams~\cite{arenas2021nfa}, and nondeterministic read-once branching
programs~\cite{meel2025nfbdd}. For a fixed alphabet and independence
relation, de Colnet, Meel, and Mathur~\cite{decolnet2026traces} give an
FPRAS and an almost-uniform sampler for the Mazurkiewicz traces
represented by accepted words of a given length. This counts words
modulo swaps of adjacent independent letters.
The NFA running-time bounds recalled in the introduction treat the
alphabet size as constant. The grammar and tree-automaton models use
explicit alphabets, while decision diagrams use Boolean variables.

\myparagraph{Counting over succinct alphabets}
Arenas et al.~\cite[Theorem~1.4]{arenas2021tree} also give an FPRAS
and a fully polynomial-time almost-uniform sampler for \emph{succinct
NFAs}, allowing exponentially large alphabets and succinct transition
labels. Their interface provides the same three kinds of oracle access
as ours. We build on the more recent dependent-sampling algorithm of
Meel and de Colnet~\cite{decolnet2025practical}, extending it to symbolic
guards. This requires handling sampled letters, overlapping guards,
and oracle errors. We give explicit query bounds and recover their
running-time bound for a constant-size explicit alphabet.

\myparagraph{Determinization and string model counting}
The \textsc{ABC} tool~\cite{aydin2015automata} counts solutions to string
constraints through path counting in deterministic automata, while
\textsc{SMC}~\cite{luu2014smc} uses generating functions to compute
bounds on solution counts. For \cntSNFA{}, determinization yields
disjoint outgoing guards whose cardinalities serve as weights for
dynamic programming or matrix powering. Symbolic
determinization can form predicate combinations locally, without
computing all global minterms~\cite{veanes2010rex}. This can be effective
when the resulting automaton is small and its guards are easy to count.
In the worst case, the number of states grows exponentially, and
counting the new guards requires quantitative access to Boolean
combinations of the input predicates. Our algorithm works directly on
the symbolic NFA and queries only its original guards.

\myparagraph{Model counting}
Approximate model counting has a long history, from the FPRAS for DNF
formulas~\cite{karp1989montecarlo} to hashing-based SAT counters such as
\textsc{ApproxMC}~\cite{chakraborty2013scalable}, which provide accuracy
and confidence guarantees through SAT-solver calls. Chistikov et
al.~\cite{chistikov2015approximate} extend solver-based approximate
counting to SMT, and Chakraborty et al.~\cite{chakraborty2016approx}
develop word-level hashing for bit-vector constraints. Model counters
and samplers can supply our guard operations, while our algorithm
combines them to count distinct accepted words. Their accuracy and
sampling guarantees must satisfy our interface, and their costs remain
explicit in our bounds.

%!TEX root=./main.tex

\section{Conclusion}
\seclabel{conclusion}

We extend the dependent-sampling FPRAS for \cntNFA{} to symbolic guards using membership tests, approximate counts, and almost-uniform samples, and formalize the core probabilistic correctness argument in Lean 4. The running time depends on the oracle costs. Sampling requires independent draws with per-letter multiplicative guarantees, and every guard must have finite denotation.

Two directions remain. Can our dependent samples yield a polynomial-time almost-uniform generator? Such generation is known for succinct NFAs~\cite{arenas2021tree}, but using our samples requires controlling dependencies and conditioning. Can the oracle tolerance $\dtol \in O(\epsl/n^2)$ be relaxed? A sharper analysis or sampling procedure could reduce the accuracy required of guard counters and samplers.

%!TEX root=./main.tex

\section*{Disclosure of Use of AI in Research}
The authors developed the research ideas, results, and mathematical proofs
presented in this paper. We then sought to use AI as much as possible for writing of TeX and the Lean formalization. 
As was expected, LLMs came up with all kinds of weird terminology and the authors focused on making the paper understandable to humans. The authors reviewed the Lean theorem statements and proof structure.

\section*{Data Availability Statement}
The Lean 4 formalization accompanying this paper is publicly
available~\cite{mathur2026symbolicartifact}.

\bibliographystyle{ACM-Reference-Format}
\bibliography{references}

\end{document}